\documentclass[11pt,final]{amsart}
\usepackage{amsfonts,amsmath,amssymb,mathtools}
\usepackage[utf8]{inputenc}
\usepackage{graphicx}
\usepackage[comma, authoryear]{natbib}
\usepackage[colorlinks,citecolor=blue]{hyperref}
\usepackage{pdflscape}
\usepackage{xspace,colortbl,multirow}
\usepackage{color}

\usepackage{outlines}
\usepackage{tikz}
\usepackage{ulem}
\usepackage{setspace}
\usepackage[color,notref,notcite]{showkeys}
\usepackage{enumerate}
\definecolor{labelkey}{rgb}{0.6,0,1}

\def\disp{\displaystyle}

\def\setl{\{1,\ldots,\ell\}}

\def\setn{\{1,\ldots,n\}}

\DeclareMathOperator*{\argmax}{arg\,max}

\newcommand{\dD}{\ensuremath{\mathbb{D}}}
\newcommand{\dE}{\ensuremath{\mathbb{E}}}

\newcommand{\I}{\ensuremath{\mathbb{I}}}

\newcommand{\dN}{\ensuremath{\mathbb{N}}}

\newcommand{\dP}{\ensuremath{\mathbb{P}}}

\newcommand{\dR}{\ensuremath{\mathbb{R}}}

\newcommand{\cA}{\ensuremath{\mathcal{A}}}
\newcommand{\cB}{\ensuremath{\mathcal{B}}}

\newcommand{\cF}{\ensuremath{\mathcal{F}}}
\newcommand{\cG}{\ensuremath{\mathcal{G}}}
\newcommand{\cH}{\ensuremath{\mathcal{H}}}

\newcommand{\cL}{\ensuremath{\mathcal{L}}}

\newcommand{\cP}{\ensuremath{\mathcal{P}}}

\newcommand{\cQ}{\ensuremath{\mathcal{Q}}}
\newcommand{\cR}{\ensuremath{\mathcal{R}}}
\newcommand{\cS}{\ensuremath{\mathcal{S}}}

\newcommand{\fL}{\ensuremath{\mathfrak{L}}}

\def\sn{n^{1/2}}

\def\bq{\mathbf{q}}

\def\bx{\mathbf{x}}
\def\by{\mathbf{y}}
\def\bz{\mathbf{z}}

\def\bQ{\mathbf{Q}}

\def\bV{\mathbf{V}}

\def\bX{\mathbf{X}}
\def\bY{\mathbf{Y}}
\def\bZ{\mathbf{Z}}

\def\balpha{{\boldsymbol\alpha}}
\def\bbbeta{{\boldsymbol\beta}}

\def\bkappa{{\boldsymbol\kappa}}

\def\bphi{{\boldsymbol\phi}}

\def\btheta{{\boldsymbol\theta}}

\def\btau{{\boldsymbol\tau}}
\def\bvarphi{{\boldsymbol\varphi}}

\newtheorem{theorem}{Theorem}

\newtheorem{prop}[theorem]{Proposition}

\newtheorem{rem}{Remark}[section]

\newtheorem{algo}{Algorithm}

\title[Goodness-of-fit tests for Hidden Markov Models?]{Are Information criteria good enough to choose the right the number of regimes in Hidden Markov Models?}

\author{Bouchra R. Nasri}
\address{Département de médecine sociale et préventive, École de santé publique, Université de Montréal, C.P. 6128, succursale Centre-ville
Montréal (Québec)  H3C 3J7 }
\email{bouchra.nasri@umontreal.ca}

\author{Bruno N. R\'emillard}
\address{Department of Decision Sciences, HEC Montr\'eal,
\newline Montr\'eal, Qu\'ebec, Canada  H3T 2A4}
\email{bruno.remillard@hec.ca}

\author{Mamadou Y. Thioub}
\address{Department of Decision Sciences, HEC Montr\'eal,
\newline Montr\'eal, Qu\'ebec, Canada  H3T 2A4}
\email{mamadou-yamar.thioub@hec.ca}

\begin{document}

\begin{abstract}
Selecting the number of regimes in Hidden Markov models is an important problem. Many criteria are used to do so, such as Akaike information criterion (AIC), Bayesian information criterion (BIC), integrated completed likelihood (ICL), deviance information criterion (DIC), and Watanabe-Akaike information criterion (WAIC), to name a few. In this article, we introduced goodness-of-fit tests for general Hidden Markov models with covariates, where the distribution of the observations is arbitrary, i.e., continuous, discrete, or a mixture of both. A selection procedure is proposed based on this goodness-of-fit test. The main aim of this article is to compare the classical information criterion with the new criterion when the outcome is either continuous, discrete or zero-inflated.  Numerical experiments assess the finite sample performance of the goodness-of-fit tests, and comparisons between the different criteria are made.


\end{abstract}

\maketitle

\section{Introduction}\label{sec:intro}

Hidden Markov models (HMM) are used in many fields to model the dynamics of complex systems such as the incidence of a specific infectious disease  \citep{Douwes-Schultz/Schmidt:2022},  animal movements \citep{Bastille-Rousseau/Potts/Yackulic/Frair/Ellington/Blake:2016, Pohle/Langrock/vanBeest/Schmidt:2017,Leos-Barajas/Photopoulou/Langrock/Patterson/Watanabe/Murgatroyd/Papastamatiou:2017}, financial assets \citep{Hamilton:1989,Hamilton:1990, Hardy:2001}, or the dynamics of biological sequences \citep{Yoon:2009}. Furthermore, \cite{Diebold/Inoue:2001} also showed that HMM can approximate long memory processes. Finally, note that hidden semi-Markov models can be seen or can be approximated by HMM with an enlarged number of regimes \citep{Zucchini/MacDonald/Langrock:2016}. Since the regimes often have an interpretation, selecting the correct number of regimes is important. Mathematically,  a hidden Markov model (HMM) depends on two different sources of randomness: a finite number of (non-observable) regimes whose dynamics are modelled by a homogeneous (stationary) or inhomogeneous (non-stationary) Markov chain and observations whose distribution depends on the regimes and possibly previous observations and covariates.
There are two important steps in implementing an HMM: (i) estimation of parameters and (ii) selection of the number of regimes. Since the regimes are not observable, the parameters can be estimated by maximizing the likelihood or using the Expectation-Maximization (EM) algorithm \citep{Dempster/Laird/Rubin:1977}. One can also use Bayesian methods as in \cite{Scott:2002}; see also \cite{Ryden:2008, Zucchini/MacDonald/Langrock:2016} for implementation issues. For the selection of the number of regimes, in many cases, researchers assumed that the number of regimes. For example, in \cite{Hardy:2001}, it is assumed that monthly returns were modelled by two-regime HMM, which makes sense since monthly returns are often almost independent. In other cases, it is difficult, if not impossible, to know or to pre-define the correct number of regimes, which makes selection methods more attractive.
Often, methods based on information criteria such as AIC \citep{Akaike:1974} and BIC \citep{Schwarz:1978} are used, which has been proved to work for some specific Hidden Markov models with a discrete finite state space for the outcomes \citep[Chapter 15]{Cappe/Moulines/Ryden:2005}.
However, this hypothesis is not met in this work since we consider outcomes with infinite support. Another information criterion is the ICL (Integrated Completed Likelihood), introduced in \cite{Biernacki/Celeux/Govaert:2000}. The ICL is very similar to the BIC, but instead of considering only the (partial) log-likelihood of the outcomes, the hidden states are estimated and plugged into the complete log-likelihood.
One can also use Bayesian-based criteria like the deviance information criterion (DIC)  \citep{Spiegelhalter/Best/Carlin/vanderLinde:2002} and the Watanabe-Akaike information criterion (WAIC) introduced by \cite{Watanabe:2010}. For an interesting review of criteria for model selection, see also \cite{Gelman/Hwang/Vehtari:2014} and references therein. Finally, for HMMs, \cite{Scott:2002} proposed to compute the posterior probability of having $s$ regimes, and the number of regimes is the most probable one according to this posterior distribution. In the literature, there have been few comparisons between information-based regime selection methods \citep{Pohle/Langrock/vanBeest/Schmidt:2017, Celeux/Durand:2008}. The comparisons are generally performed when there are no covariates, and the distribution is either continuous or discrete.
A common characteristic of these selection methods is that the maximum number of regimes  $\ell_{max}$ should be chosen in advance, and the information criterion must be computed for all regimes $\ell\le \ell_{max}$. Furthermore, choosing the number of regimes $\ell^\star$ minimizing a given information criterion does not imply that the model is correct.\\

To assess the fit of the selected model, \cite{Zucchini/MacDonald/Langrock:2016} and \cite{Pohle/Langrock/vanBeest/Schmidt:2017} suggest plotting what they called ``pseudo-residuals''. However, they did not propose formal goodness-of-fit tests.
Recently, \cite{Nasri/Remillard:2019a, Nasri/Remillard/Thioub:2020} proposed formal goodness-of-fit tests based on pseudo-observations constructed from the conditional Rosenblatt's transforms for general dynamic models with continuous outcomes (including multivariate HMMs with continuous distribution). Under the null hypothesis that the dynamic model is correct, these pseudo-observations are approximately independent and uniformly distributed. \cite{Nasri/Remillard:2019a, Nasri/Remillard/Thioub:2020} also proposed to select the regimes based on goodness-of-fit tests.
  Goodness-of-fit tests based on pseudo-observations are not new and have been proposed in univariate settings by \cite{Diebold/Gunther/Tay:1998} and \cite{Bai:2003}, when the associated cdfs are continuous.
    The problem with this technique is that when the cdfs are not continuous, the pseudo-observations are no longer approximately independent or uniformly distributed. To overcome this problem, randomization methods were proposed in literature \citep{Brockwell:2007, Kheifets/Velasco:2013}. Note that recently, \cite{Kheifets/Velasco:2017} proposed another method not involving randomization for purely discrete time series, which can be generalized for mixtures of discrete and continuous distributions (arbitrary data).

The main aim of this paper is to compare the selection of the number of regimes using information criteria to the selection method based on goodness-of-fit tests. Here, we consider general HMMs with covariates, including possibly past observations, and where the distribution of the observations is continuous, discrete or a mixture of both. These models are described in Section \ref{sec:hmm}, together with an estimation method based on the EM algorithm, a generalized version of goodness-of-fit tests, and the proposed selection procedure for the number of regimes.
In Section \ref{sec:num},  under different data-generating processes, numerical experiments are performed to assess the power of the goodness-of-fit tests, as well as the comparison of the precision of the selection methods for the number of regimes. The generalized goodness-of fit procedure is compared to classical selection methods used in the literature.
Finally, in Section \ref{sec:ex}, we present an application of the proposed methodologies for the incidence of COVID-19 cases in counties of New York state.

\section{Description of the models and estimation of parameters}\label{sec:hmm}

Suppose that  $\tau_t \in \setl$ is the regime at time $t\in \setn$, $Y_t$ is the variable of interest, and $\bZ_t$ is the vector of covariates. Let $\cF_{t-1}$ be the information (sigma-algebra) generated by $\{ Y_1,\ldots,Y_{t-1},\bZ_{1}, \ldots, \bZ_t\}$ and let $\cG_{t-1}$ be the information generated by $\{\tau_1,\ldots,\tau_{t-1}\}$ and $\cF_{t-1}$. Further set
$\bX_t = (Y_{t-1},\ldots,Y_{t-p},\bZ_{t})$, $t\ge 1$.

Hidden Markov  models considered in this paper have the following components: a series of transition matrices  $\bQ_t = \bq(\bkappa,\bx_t)$, with parameter $\bkappa=(\bkappa_1,\ldots,\bkappa_\ell) \in \dR^{\ell\times \ell}$, where
$ \disp Q_{t,j,k} = q_{jk}(\bkappa_j,\bx_t) = P(\tau_t=k|\cG_{t-1},\tau_{t-1}=j)$, $j,k\in \setl$,
and cumulative probability distributions
$\disp G_{j,\bbbeta_j}(y,\bx_t) = P(Y_t\le y|\cG_{t-1},\tau_t=j,\bX_t=\bx_t)$, $ \bbbeta_j\in\cB_j$,
where $\cB_j$ is a parameter space.
It is assumed that for any $j\in\setl$, $G_{j,\bbbeta_j}$ has density
$g_{j,\bbbeta_j}(y_t,\bx_t)$ with respect to a reference measure $\fL$. The parameter of interest is $\btheta = (\bbbeta,\bkappa)$, with $\bbbeta = (\bbbeta_1,\ldots,\bbbeta_\ell)\in \cB_1\times \cdots\times \cB_\ell$ and $\bkappa = (\bkappa_1,\ldots,\bkappa_\ell) \in \dR^{\ell\times \ell}$.  In this article, references measures are of the form $\fL = \delta_{\dN^\star}+\cL$, where $\delta_\cA$ is the counting measure on a countable set $\cA$, and $\cL$ is Lebesgue's measure. This way, one can deal in a unified way with arbitrary distributions, i.e., continuous, discrete, as well as mixtures of discrete and continuous distributions. Next, let $\delta$ be the counting measure on $\setl$ and assume that $\eta_0$ is the distribution of $\tau_0$. Most of the time, $\eta_0$ has no effect on the estimation of the parameters, particularly if the Markov chain is ergodic, meaning that one can start with another $\eta_0$, as long as $\eta_0(j)>0$ for all $j\in\setl$.  With respect to reference measure $\delta^{\otimes n} \times \fL^{\otimes n}$,
the joint conditional density of $\btau=(\tau_1,\ldots,\tau_n)$ and $\bY=(Y_1,\ldots,Y_n)=\by$ given $\bZ = (\bZ_1,\ldots,\bZ_{n})=\bz$ and $\tau_0$ is
\begin{equation}\label{eq:dens-joint}
f_\btheta(\btau,\by,\bz) = \eta_0(\tau_0)\left(\prod_{t=1}^{n} Q_{t,\tau_{t-1},\tau_{t}}\right) \times \prod_{t=1}^n g_{\tau_t,\bbbeta_{\tau_t} }(y_t, \bx_t).
\end{equation}

\subsection{Estimation}

Since most regimes are not observable, one can use the EM algorithm \citep{Dempster/Laird/Rubin:1977}, which proceeds in two steps. In the first step (expectation or E-step), one computes  $\cQ_{\by,\bz}\left(\tilde\btheta,\btheta\right)  =  \dE_\btheta\{\log f_{\tilde\btheta}(\tau,\bY,\bZ) |\bY=\by,\bZ=\bz\}$.
From Appendix \ref{app:genEM},
\begin{equation}\label{eq:estep}
   \cQ_{\by,\bz}(\tilde\btheta,\btheta) =  \log{\eta_0(\tau_0)}+ \sum_{t=1}^{n} \sum_{j=1}^\ell\sum_{k=1}^\ell \Lambda_{t,\btheta}(j,k)\log\tilde Q_{t,jk}  + \sum_{t=1}^n \sum_{j=1}^\ell  \lambda_{t,\btheta}(j)  \log g_{j,\tilde\bbbeta_j}\left(y_t,\bx_t\right),\\
\end{equation}
where $
\lambda_{t,\btheta}(j) = \dP_\btheta(\tau_t=j|\bY=\by,\bZ=\bz)$ and $\Lambda_{t,\btheta}(j,k)=  \dP_\btheta(\tau_{t-1}=j,\tau_{t}=k|\bY=\by,\bZ=\bz)$,
for all $t\in \setn$ and $j,k \in \setl$.
In the second step,  (maximisation or M-step), one computes
$\disp \btheta^{(k+1) } = \argmax_{\btheta} \cQ_y\left(\btheta,\btheta^{(k)}\right)$,
starting from an initial value $\btheta^{(0)}$. One can see from \eqref{eq:estep} that parameters $(\bbbeta_1, \ldots, \bbbeta_\ell)$ and $(\bkappa_1,\ldots,\bkappa_\ell)$  are estimated separately for each regime.
As $k\to\infty$, $\btheta^{(k)}$ converges to the maximum likelihood estimator of the density of $\bY$.
The formulas to implement the EM algorithm are given in 
Appendix \ref{app:genEM}.

\subsection{Goodness-of-fit tests}\label{ssec:gof}

For our models,  the conditional distribution function  $F_t$ of $Y_t$ given $\cF_{t-1}$ is expressed as a mixture. More precisely,
if $\eta_{t,\btheta}(j) = \dP_\btheta(\tau_t = j | Y_1 =y_{1},\ldots,Y_t=y_{t}, \bZ_1=\bz_1,\ldots, \bZ_t=\bz_t)$,
$t\in \setn$, then
\begin{equation}\label{eq:cond_cdf}
F_{t,\btheta}(y) = P_\btheta(Y_t\le y| \cF_{t-1}) = \sum_{k=1}^\ell G_{j,\bbbeta_j}(y,\bx_t)W_{t-1}(j) , \text{with} \quad W_{t-1}(j)=\sum_{k=1}^\ell \eta_{t-1,\btheta}(k) Q_{t,kj}.
\end{equation}
A recursive formula for computing $\eta_{t,\btheta} $ is given by Equation \eqref{eq:eta}
in Appendix \ref{app:genEM}.
It follows from \eqref{eq:cond_cdf} that the conditional density $f_t$ of $Y_t$ given $\cF_{t-1}$ is
\begin{equation}\label{eq:cond_pdf}
f_{t,\btheta}(y) = \sum_{k=1}^\ell g_{j,\bbbeta_j}(y,\bx_t)W_{t-1}(j).
\end{equation}
Equation \eqref{eq:cond_pdf} can in turn be used to compute the log-likelihood as well as AIC and BIC values.
Next, if each $G_{j,\bbbeta_j}$ is continuous for all $j\in\setl$, then $U_t = F_{t,\btheta}(Y_t)$ are iid and uniformly distributed over $(0,1)$, while this is no longer true if one of the $F_j$s has discontinuities, as in the zero-inflated case. In the latter case, one can construct
$$
U_t = (1-V_t)F_{t,\btheta}(Y_t-)+ V_t F_{t,\btheta}(Y_t) = F_{t,\btheta}(Y_t-)+V_t \Delta F_{t,\btheta}(Y_t),
$$
where $V_1,\ldots, V_n$ are iid uniform independent of $Y_1,\ldots,Y_n, \bZ_1,\ldots,\bZ_n$, $F_{t,\btheta}(y-) =  P(Y_t < y| \cF_{t-1})$, and $\Delta F_{t,\btheta}(y) =  P(Y_t = y| \cF_{t-1})$. In this case, under the null hypothesis that the model is correct, it follows from \cite{Brockwell:2007} that the $U_t$s are iid and uniformly distributed over $(0,1)$.
This result is not new. It was already used in the context of randomized tests for iid observations by \citet[Section 5.3]{Ferguson:1967}, and it appeared in the proof of Lemma 3.2 in \cite{Moore/Spruill:1975}.
Note that in the continuous case, one recovers $U_t = F_{t,\btheta}(Y_t)$, so there is no added randomness.
For a given number of regimes, the null hypothesis can be tested using the empirical cdf $D_n$ of the  pseudo-observations $u_{n,t}= (1-v_t)F_{t,\btheta_n}(y_t-)+ v_t F_{t,\btheta_n}(y_t) $, $t\in \setn$, defined for any $u \in [0,1]$, by
 $\disp
 D_{n}(u)= \dfrac{1}{n} \sum_{t=1}^{n} \mathbf{1}\{u_{n,t} \leq u\}$.
 As shown in \cite{Bai:2003} and \cite{Remillard:2011a}, the empirical process $\dD_n$ defined by $\dD_n(u) = \sn\{D_n(u)-u\}$, $u\in [0,1]$, converges in an appropriate space to a continuous centred Gaussian process $\dD$.
  As a test statistic, one could use the Cram\'er-von Mises statistic $S_n$ defined  by
\begin{eqnarray}\label{eq:cvm}
S_n & = & \cS_{n}{\left( u_{n,1},\ldots,u_{n,n} \right)} = \int_0^1 \dD_n^2(u) du  = \sum_{t=1}^{n} \left\{u_{n:t}-\frac{(t-1/2)}{n} \right\}^{2} + \frac{1}{12n},
\end{eqnarray}
where  and $u_{n:1} < \dots < u_{n:n}$ are the ordered values. Since $S_n$ is a measure of distance between the empirical distribution and the uniform distribution, large values of $S_n$ lead to the rejection of the null hypothesis.
One could also use the Kolmogorov-Smirnov statistic $T_n$ defined  by
\begin{equation}\label{eq:ks}
T_n  = \sup_{u\in [0,1]} |\dD_{n}{(u)}| = \sn \max_{1\le t\le n}\max\left\{\left|u_{n:t}-\dfrac{t}{n}\right|, \left|u_{n:t}-\dfrac{t-1}{n}\right|\right\}.
\end{equation}
Since $T_n$ is also a measure of distance between the empirical cdf and the uniform distribution, large values of $T_n$ lead to the rejection of the null hypothesis.

The only problem is that the limiting distribution of the empirical process $\dD_n$ and the test statistics $S_n$ and $T_n$  depend on the estimation error of the unknown parameter $\btheta$. To overcome this difficulty, a  parametric bootstrap method, described in Algorithm \ref{algo1}, is used to estimate the $P$-values.
The validity of the parametric bootstrap procedure was proven in \cite[Proposition 1]{Remillard:2011a} for general dynamic models, including HMMs. The algorithm is described for $S_n$ but it can be used for other statistics as well.

\begin{algo}\label{algo1}
For a given number of regimes $\ell$,  let $\btheta_n$ be the maximum likelihood estimator of $\btheta$ obtained from the EM algorithm. Then compute the statistic $ S_n = \cS_{n}{\left( u_{n,1},\ldots,u_{n,n} \right)}$ given be \eqref{eq:cvm}.   Next, for $k=1,\ldots, B$, $B$ large enough, repeat the following steps:
\begin{enumerate}
\item Generate a random sample $Y_{1}^*,\ldots,Y_{n}^*$ with values $y_{1}^*,\ldots,y_{n}^*$ from distribution $\textbf{P}_{\btheta_n}$, i.e., from an HMM with parameter $\btheta_n$, and generate uniform iid $V_1^*,\ldots, V_n^*$  with  values $v_1^*,\ldots, v_n^*$.\\

\item Compute the estimator $\btheta_n^*$ from $Y_{1}^*,\ldots,Y_{n}^*$.\\

\item Compute the pseudo-observations $u_{n,t}^* =  \left(1-v_t^*\right) F_{t,\btheta_n^*}\left(y_t^*-,\bx_t\right) + v_t^*  F_{t,\btheta_n^*}\left(y_t^*,\bx_t\right)$,  $t\in\setn$, using Equation \eqref{eq:cond_cdf}, and compute $S_n^{(k)} =
\cS_{n}{\left( u_{n,1}^*,\ldots,u_{n,n}^* \right)}$.
\end{enumerate}
Then, an approximate $P$-value for  $S_n$ is given by
$\disp
\frac{1}{B} \sum_{k=1}^{B} \mathrm{1} \left(S_{n}^{(k)}\ge S_{n} \right)$.
\end{algo}
One can also implement the average-randomisation procedure  \citep{Machado/SantosSilva:2005} to improve the power of the proposed tests $S_n$ or $T_n$. To this end, one simulates $M$ independent samples $\bV_k=(V_{k1},\ldots,V_{kn})$, $k\in\{1,\ldots,M\}$ and compute associated empirical processes $\dD_n^{(1)},\ldots, \dD_n^{(M)}$. Then, set
$\disp
\bar \dD_{n,M} = \frac{1}{M}\sum_{k=1}^M \dD_n^{(k)}$,
and compute the associated Cram\'er-von Mises statistic $\bar S_{n,M} = \int_0^1 \bar\dD_{n,M}^2(u)du$. An explicit expression for $\bar S_{n,M}$ is given by Equation \eqref{eq:Snbar} in Appendix \ref{app:average-Sn}.
The approximate $P$-value can then be computed using a slight modification of Algorithm \ref{algo1} by using averaging.
It was shown in  \cite{Kheifets/Velasco:2017} that this procedure, using as few as $M=25$ samples, was almost as efficient as their proposed methodology for purely discrete data.

\begin{figure}[!ht]
{\centering
	\includegraphics[scale = 0.45]{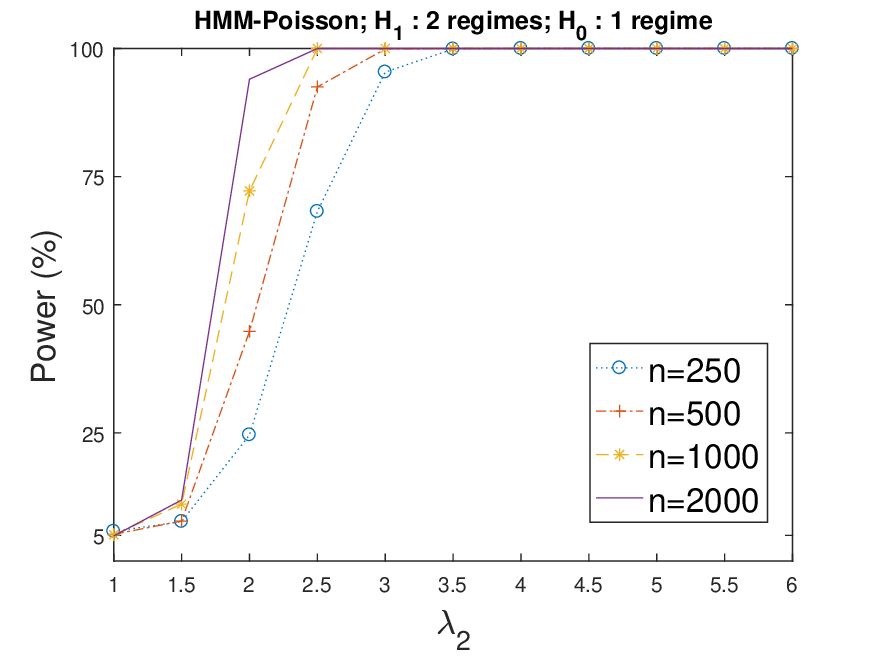}
	\includegraphics[scale = 0.45]{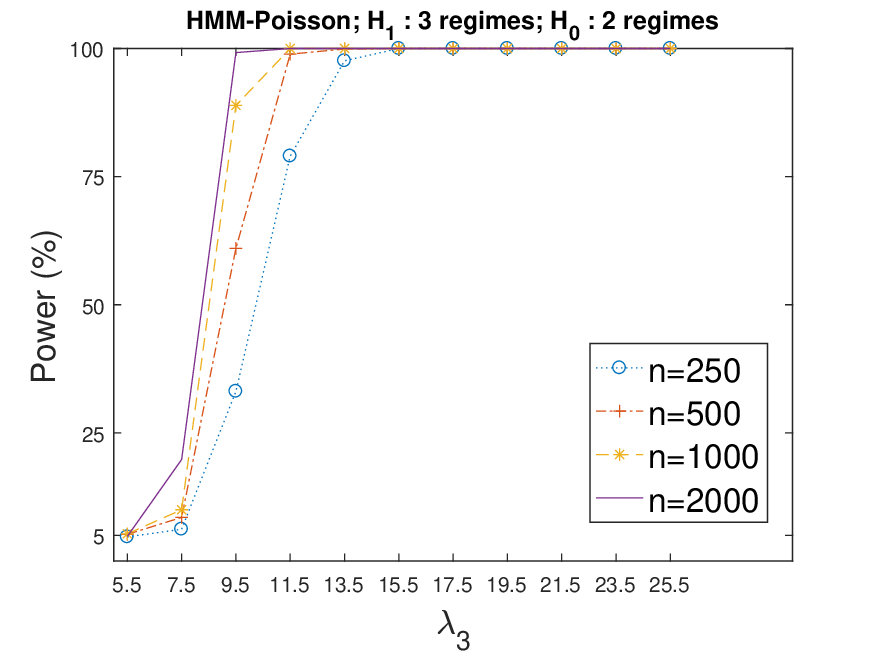}
    \caption{Power curves for Poisson regime-switching models.}
	\label{fig:powerPoisson}
}
\end{figure}

Although the proposed goodness-of-fit tests are generally consistent, it seems that for Poisson models, the mean of the regimes must be quite different in order to detect more regimes, especially if the sample size is small.  This is illustrated in Figure \ref{fig:powerPoisson}, where the power, for $n\in\{250,500,1000,2000\}$  is displayed as a function of the constant means $\lambda_1=1$, $\lambda_2 \in [1,6]$ for the null hypothesis of 1 regime, and $\lambda_1=1, \lambda_2=5$, $\lambda_3 \in [5,25]$ for the null hypothesis of 2 regimes.

\subsection{Selection procedure for the number of regimes}\label{ssec:selection}

As proposed in \cite{Nasri/Boucher/Perreault/Remillard/Huard/Nicault:2020, Nasri/Remillard/Thioub:2020},  at the 5\% level, we choose the number of regimes  $\ell^*$ as the first $\ell$ for which the $P$-value of the goodness-of-fit test is larger than $5\%$, starting with $1$ regime, unless the model is a zero-inflated one, in which case one starts with $2$ regimes. This selection procedure was shown  to be quite effective for determining the number of regimes in HMM copulas \citep{Nasri/Remillard/Thioub:2020}. Of course, one could choose a level different from 5\%.\\

This methodology uses multiple testing. However, if a goodness-of-fit test is powerful enough, multiple testing has no significant effect. For example, if $\cP_1$ and $\cP_2$ are $P$-values
computed for the null hypothesis that model $i\in \{1,2\}$ is correct, and if model 2 is the correct one, then the probability of choosing model 2 over model 1 is
$\disp P(\{\cP_2 > 5\%\}  \cap \{ \cP_1 < 5\%\}) \approx  P( \cP_2 > 5\% | \cP_1 < 5\%) \approx 95\%$,
if the test is powerful enough,
meaning that $P(\cP_1< 5\%)$ is close to 1. These computations extends to any finite number of
models, as long as the sample size is large enough. Therefore, for the proposed selection of the correct number of regimes to work, all one needs is that the goodness-of-fit test  will reject the null hypothesis for $\ell$ regimes with large probability, whenever $\ell< \ell^{true}$ since in this case,
$$
P\left( \cup_{\ell <\ell^{true}} \{\cP_l \ge 5\%\}\right) \le \sum_{\ell=1}^{\ell^{true}-1}P\left( \cP_l \ge 5\% \right)  \approx 0.
$$
As a result, the probability of selecting $\ell^\star=\ell^{true}$ should be close to $95\%$.

\subsection{Other selection methods}\label{ssec:information}

Having defined our proposed selection method, we now define the other selection methods we will consider for the numerical comparisons.

The main information-based criteria are AIC, BIC, and ICL, defined respectively by
$\disp \rm{AIC} = 2 \times {\rm param} -2\sum_{t=1}^n \log f_t(y_t)$,
$\disp \rm{BIC} = \log{n}\times {\rm param} -2\sum_{t=1}^n \log f_t(y_t)$,
and $\disp
\rm{ICL} = \log{n}\times {\rm param} -2\sum_{t=1}^n \log g_{\hat \tau_t,\bbbeta_{\hat \tau_t}}(y_t;\bx_t)-2\sum_{t=1}^n \log{Q_t\left(\hat \tau_{t-1},\hat \tau_t\right)}$,
where ${\rm param}$ is the total number of estimated parameters, and $\hat \tau_t$ is the most probable state at time $t\in\setn$.
In \cite{Pohle/Langrock/vanBeest/Schmidt:2017}, the authors considered two-state HMM with two very different distributions and transition matrix defined with probability $0.9$ of staying in the same state. They considered seven scenarios ($H_1$) which were misspecifications of the null hypothesis. Their conclusion was that the ICL criterion was the best. They also concluded that for these scenarios, the AIC and BIC had a tendency to overestimate the number of regimes.  We will see next that we get opposite results.\\

Next, we present Bayesian selection criteria.
Suppose the log-density of the observations is $\log{f(\by|\bvarphi) }= \sum_{t=1}^n \log{f_t(y_t|\bvarphi)}$, and set $D(\bvarphi) = -2\log{f(\by|\bvarphi)}$.
If $\left\{\bvarphi^{(k)}; k=1,\ldots,N_B\right\}$ is a sample from the posterior distribution, set
 $\bar \bvarphi = \dfrac{1}{N_B}\sum_{k=1}^{N_B}\bvarphi^{(k)}$,  and let
${\rm var}_{\bvarphi}\left\{   \log f_t\left(y_t|\bvarphi\right)    \right\}$ be the sample variance of the $\log f_t\left(y_t|\bvarphi^{(k)}\right)$, $k=1,\ldots,N_B$.
Further set  $p_D = \bar D-D(\bar\bvarphi)$, where
$\bar D = \frac{1}{N_B}\sum_{k=1}^{N_B} D\left(\bvarphi^{(k)}\right)$.
The Deviance Information Criterion (DIC)  \citep{Spiegelhalter/Best/Carlin/vanderLinde:2002} is
$ \rm{DIC} = \bar D + p_D = 2p_D+D(\bar\bvarphi)$, and the Watanabe-Akaike information criterion (WAIC) \citep{Watanabe:2010} is
 $\disp
    \rm{WAIC} = -2 \left[ \sum_{t=1}^{n} \log \left(\frac{1}{S} \sum_{k=1}^{N_B} f_t(y_t|\bvarphi^{(k)})\right)- \sum_{t=1}^{n} {\rm var}_\bvarphi\left\{ \log f_t\left(y_t|\bvarphi\right)\right\}\right]$.
For an interesting review of criteria for model selection, see also \cite{Gelman/Hwang/Vehtari:2014} and reference therein.
Finally, for HMMs, \cite{Scott:2002} proposed to compute the posterior probability of having $s$ regimes. If $\gamma_S$ is the a priori distribution of the number of regimes $S\in \{1,\ldots, s_{max}\}$, then
$
P(S=s|\by,\bz)  =  \frac{\gamma_S(s) \sum_{k=1}^{N_B} f\left(\by,\bz|\bvarphi_s^{(k)} \right)}{\sum_{j=1}^{s_{max}} \sum_{k=1}^{N_B} \gamma_S(j) f\left(\by,\bz|\bvarphi_j^{(k)} \right)}$,
where  $\left\{\bvarphi_s^{(k)}; k=1,\ldots,N_B\right\}$ is a sample from the posterior distribution assuming there are $s$ regimes, $s \in \{1,\ldots, s_{max}\}$.

\section{Numerical experiments}\label{sec:num}
In this section we perform Monte Carlo experiments for: (i) assessing the power of the proposed goodness-of-fit tests; (ii) comparing the regime selection procedures proposed in Section \ref{ssec:selection}.
To this end, we generated random samples of size $n\in\{250,500\}$ from several ARX-HMM models described in Section \ref{ssec:dgp}. For each  sample size and for each model, and for a given number of regimes, we perform 1000 replications and in each replication, whenever needed, $B=100$ bootstrap samples  were used to compute the $P$-value of tests statistic $ \bar S_{n,1}$,  $\bar S_{n,25}$, and  $\bar S_{n,50}$.
In practice, when there are no replications, larger bootstrap samples should be used.
In Section \ref{ssec:power}, we assess the power of the proposed goodness-of-fit tests for different DGPs.  Next, in Section \ref{ssec:selectionExp}, we compare the seven regime selection procedures for the same DGPs, as well as for a Poisson model with constant means in each regime. In each case, we used $N=100$ replications. To reduce the computational burden, only a small number of  regimes were considered. However, in practice, larger number are necessary, e.g., for daily returns of assets (4 regimes for 2-year sequences and up to 6 for longer series).

\subsection{Data generating processes (DGP)}\label{ssec:dgp}

For the numerical experiments, we considered autoregressive HMM with Gaussian and Poisson distributions with covariates described next. They all satisfy Equation \eqref{eq:dens-joint}.
We also considered their zero-inflated counterparts.

\begin{itemize}
    \item[($M_1$)]  Gaussian ARX-HMM: In this model, the conditional distribution of $Y_t$ given $\cF_{t-1}$ and $\tau_t=j$ is  Gaussian with mean $\mu_{tj} =  \sum_{k=1}^p \phi_{jk}Y_{t-k}+ \balpha_j^\top \bz_t$ and standard deviation  $\sigma_j$. In this case, $\bx_t = ( y_{t-1},\ldots, y_{t-p}, z_{t,1},\ldots, z_{t,r})^\top$. The case $p=1$ and covariate $z_t \equiv 1$ was considered in \cite{Ang/Bekaert:2002, Caccia/Remillard:2018}, while the general case appeared in \cite{Fruhwirth-Schnatter:2006}. It is  assumed that $z_{t1}=1$ for all $t$. Here, one assumes that for any $j\in\setl$, all zeros of the polynomial
    $\Phi_j(\zeta) =1-\sum_{k=1}^p \phi_{jk}\zeta^k$ are outside the unit circle. Also $\bbbeta_j = (\bphi_j,\balpha_j,\sigma_j)^\top$, $j\in\setl$.\\

    \item[($M_2$)] Poisson ARX-HMM : This model is similar to the one proposed in \cite{Kedem/Fokianos:2002,Fokianos/Tjostheim:2011}, where $g_{j,\bbbeta_j}$ is the Poisson density with parameter $\mu_{tj}$, where, in the log-linear case, $
    \mu_{tj} =\exp\left( \sum_{k=1}^p \phi_{jk}\log(1+y_{t-k})+ \balpha_j^\top \bz_t\right)$,
     with $\bbbeta_j = (\bphi_j,\balpha_j)^\top$, $j\in\setl$. Here, $z_{t,1}=1$ for all $t$, and $\bx_t = ( y_{t-1},\ldots, y_{t-p}, \bz_{t})^\top$.  When the covariates are all non negative, one can also consider the linear case, where $\mu_{tj} = \sum_{k=1}^p \phi_{jk}y_{t-k}+\balpha_j^\top \bz_t$, with the additional constraints $\phi_{jk}\ge 0$, $\sum_{k=1}^p \phi_{jk}<1$, $\alpha_{j1}>0$, and $\alpha_{jk}\ge 0$, $k\in\{2,\ldots, r\}$.  Note that contrary to \cite{Fokianos/Tjostheim:2011, Doukhan/Fokianos/Rynkiewicz:2021}, $\mu_{tj}$ does not depend on $\mu_{t-1,j}$, so there is no path dependence problem \citep{Berentsen/Bulla/Maruotti/Stove:2018}.\\

   \item[($M_3$)] Zero-inflated Gaussian ARX-HMM : In this model, the first regime is observable, corresponding to $Y_t=0$, while the other regimes are continuous as in the Gaussian ARX-HMM.\\

  \item[($M_4$)] Zero-inflated Poisson ARX-HMM : Here, the first regime is not observable, and corresponds to the density $g_1(y)=\I(y=0)$, while the other regimes are  as in the Poisson ARX-HMM. This is an extension of the case $r=2$ considered in \cite{Douwes-Schultz/Schmidt:2022}.\\
\end{itemize}
For the first set of experiments, we considered a stationary case, while  for the second set of experiments, there is linear trend. More precisely, the  means $\mu_{t,\ell}^{(j)}$ for experiment $j\in \{1,2\}$, are defined by
\begin{eqnarray*}
\mu_{t,\ell_0+1}^{(1)}  & = & 0.5Y_{t-1}  + 0.1Y_{t-2} -0.5 + 0.1 z_{t2}, \text{ (Gaussian case)},\label{eq:mu1G}\\
\mu_{t,\ell_0+1}^{(1)}  & = & 0.5Y_{t-1}  + 0.1Y_{t-2} + 2+ 0.1 z_{t2}, \text{ (Poisson case)},\label{eq:mu1}\\
\mu_{t,\ell_0+2}^{(1)}  & = & 0.3Y_{t-1} + 0.6Y_{t-2} + 1+ 0.5 z_{t2}, \label{eq:mu2}\\
\mu_{t,\ell_0+1}^{(2)}  & = & 0.5Y_{t-1} +10+5t/n , \label{eq:mu1seas}\\
\mu_{t,\ell_0+2}^{(2)}  & = & 0.75Y_{t-1} +8 + 4t/n, \label{eq:mu2seas}
\end{eqnarray*}
where  $z_{t2}\sim {\rm Exp}(1)$, $\ell_0 = 1$ for zero-inflated models, and $\ell_0 = 0$ otherwise.
The transition matrices  $\bQ_{\ell_0,\ell_1}$ used in the numerical experiments are defined as follows:
     $\bQ_{1,1} = \bQ_{0,2} = \begin{pmatrix}
0.94 & 0.06 \\
0.03 & 0.97
    \end{pmatrix}$,
  $\bQ_{1,2} =
\begin{pmatrix} 0.25000  & 0.25000  &  0.50000\\
    0.37500  &  0.58750  &  0.03750\\
    0.37500  &  0.01875  &  0.60625 \end{pmatrix}$
where $\ell_0=1$ for a zero-inflated model and $\ell_0=0$ otherwise, and $\ell_1$ is the number of non-zero regimes. 
It follows that the percentage of time regime 1 appears using $\bQ_{1,1}, \bQ_{0,2}$ or $\bQ_{1,2}$ is $33.\bar3\%$. Also, in the Gaussian cases ($M_1, M_3$), $\sigma_{\ell_0+1} = 0.8$ and $\sigma_{\ell_0+2} = 0.1$.
Finally, note that in the numerical experiments, we only used the Cram\'er-von Mises statistics $S_{n,M}$ with $M\in \{1,25,50\}$, since preliminary calculations showed that the Kolmogorov-Smirnov statistics had always a lower power.

\subsection{Level and power of the goodness-of-tests based on $S_{n,M}$}\label{ssec:power}
The results of the experiments for DGPs $M_1$ and $M_2$ are presented in Table \ref{tab:powerM12}, while those for DGPs $M_3$ and $M_4$ appear in Table \ref{tab:powerM34}. For each of these experiments, we considered three sample sizes, namely  $n\in\{100,250,500\}$. Also,
$\ell_1$ denotes the number of non-zeros regimes under $H_1$.

For a target level of 5\%, a 95\% confidence interval for the level, based on $N=1000$ replications, is $(3.65\%,6.35\%)$, or the maximum error is $1.35\%$. When the error is less than $1.35\%$, the test is said to be accurate, according to \cite{Batsidis/Martin/Pardo/Zografos:2014}, while a test with  empirical level smaller than 3.65\% would be conservative, and if the empirical level is greater than 6.35\%, it is said to be liberal.

 First, in the Gaussian case (DGP $M_1$), the results from Table \ref{tab:powerM12} indicate that the test is accurate for each sample sizes, since the levels are not significantly different from the 5\% targets. Also, the power for rejecting the null hypothesis of $1$ regime when there are $2$ regimes is very large, being $87.1\%$ when $n=100$ and $99.9\%$ when $n=250$. There was no need to consider the case $n=500$. Next, in the  Poisson case (DGP $M_2$), the tests based on $S_{n,1}$, $S_{n,25}$ and $S_{n,50}$ are  mostly accurate when $n\in\{250,500\}$, since the levels are also very close to the 5\% targets, but the tests are liberal when $n=100$. The test  is slightly conservative for  $n=500$ when using $S_{n,1}$, which is not recommended anyway. However, the power of tests are not very good, unless $n=500$. Also, the type of experiment is significant, since the results for Exp 2 are better than those for Exp 1, for all sample sizes. The conclusion from these experiments in the Poisson case is that in order to detect the correct number of regimes, either the differences of means are large or the sample size is large. This is coherent with the results illustrated in Figure \ref{fig:powerPoisson}.

Next, from the results of the  zero-inflated DGPs $M_3$ and $M_4$ displayed in Table \ref{tab:powerM34}, all tests are mostly accurate, even when $n=100$, since that the levels are not significantly different from the 5\% targets. The only exception is  for $M_4$, where the tests are mostly conservative when $n=250$. Also,  the power in the Gaussian case (DGP $M_3$) is very good, even when $n=100$. In the Poisson case (DGP $M_4$), the result are very good for rejecting $\ell=1$ regimes when there are in fact $3$ regimes, even for small sample sizes.  Note that the case $\ell=1$ corresponds to a model with no regimes, not the zero-inflated regime. In the Poisson case, the rejection of the null hypothesis of $\ell=3$ regimes when $\ell_1=1$, is not very good, but at least it improves when the sample size increases. Again testing for  and $\ell=2$ regimes when $\ell_1=2$  is not very good. Finally, one can see that the results for statistics $S_{n,25}$ and $S_{n,50}$ are very similar, and are better in general than for $S_{n,1}$, which we do not recommend.

\begin{table}[!ht]
\caption{Percentage of rejection of the null hypothesis of $\ell$ regimes at the 5\% level for DGPs $M_1$ and $M_2$ and for samples of size $n\in \{100,250,500
\}$, using $N=1000$ replications and $B=100$ parametric bootstrap samples. Under $\cH_1$, the true number of regimes is $\ell_1 \in \{1,2\}$. The empirical levels are displayed in bold. }\label{tab:powerM12}
\centering
{\tiny
\begin{tabular}{cc|ccc|ccc|ccc|ccc|ccc}
\hline
\multicolumn{11}{l}{Exp 1: AR(2) with ${\rm Exp}(1)$ covariate}\\
\hline
& & \multicolumn{3}{c|}{$M_1$, $n=100$} & \multicolumn{3}{c|}{$M_1$, $n=250$} & \multicolumn{3}{c|}{$M_2$, $n=100$} & \multicolumn{3}{c|}{$M_2$, $n=250$} & \multicolumn{3}{c}{$M_2$, $n=500$}  \\
 & & \multicolumn{3}{c|}{$\ell$} & \multicolumn{3}{c|}{$\ell$} & \multicolumn{3}{c|}{$\ell$} & \multicolumn{3}{c}{$\ell$} \\
$\ell_1$ & Statistic & $1$  & $2$ & $3$ & $1$  & $2$ & $3$  & $1$  & $2$ & $3$ & $1$  & $2$ & $3$ \\
\hline
\multirow{3}{*}{1}
&  $\bar S_{n,1}$  & \textbf{5.2} & 3.4 & 1.3 &\textbf{4.8} & 2.9 & 2.4 & \textbf{4.9} & 6.6 & 6.4 & \textbf{4.5} & 7.7 & 8.1 & \cellcolor[gray]{0.85} & \cellcolor[gray]{0.85} &\cellcolor[gray]{0.85}\\
&  $\bar S_{n,25}$ & \textbf{5.2} & 3.4 & 1.3 &\textbf{4.8} & 2.9 & 2.4 & \textbf{5.3} & 7.7 & 6.8 & \textbf{4.9} & 7.2 & 8.3 &\cellcolor[gray]{0.85} & \cellcolor[gray]{0.85} &\cellcolor[gray]{0.85} \\
&  $\bar S_{n,50}$ & \textbf{5.2} & 3.4 & 1.3 &\textbf{4.8} & 2.9 & 2.4 & \textbf{5.2} & 7.5 & 6.7 & \textbf{5.3} & 6.9 & 8.4 &\cellcolor[gray]{0.85} & \cellcolor[gray]{0.85} &\cellcolor[gray]{0.85} \\
 \hline
\multirow{3}{*}{2}
& $\bar S_{n,1}$  & 87.1 & \textbf{4.7} & 2.8 & 99.9 &\textbf{4.1} & 2.6 & 7.4 & \textbf{6.6} & 8.5& 12.2 & \textbf{5.8} &7.3 & 26.1 & \textbf{3.6} & 7.6\\
& $\bar S_{n,25}$ & 87.1 & \textbf{4.7} & 2.8 & 99.9 &\textbf{4.1} & 2.6 & 8.0 & \textbf{6.5} & 8.5& 13.3 & \textbf{6.1} &6.5 & 26.7 & \textbf{4.9} & 6.5\\
& $\bar S_{n,50}$ & 87.1 & \textbf{4.7} & 2.8 & 99.9 &\textbf{4.1} & 2.6 & 7.9 & \textbf{6.5} & 8.4& 13.4 & \textbf{6.1} &6.6 & 27.1 & \textbf{4.9} & 6.5\\
\hline
\multicolumn{11}{l}{Exp 2: AR(1) with  linear trend}\\
\hline
&  & \multicolumn{3}{c|}{$M_1$, $n=100$} & \multicolumn{3}{c|}{$M_1$, $n=250$} & \multicolumn{3}{c|}{$M_2$, $n=100$} & \multicolumn{3}{c|}{$M_2$, $n=250$} & \multicolumn{3}{c}{$M_2$, $n=500$}  \\
 && \multicolumn{3}{c|}{$\ell$} & \multicolumn{3}{c|}{$\ell$} & \multicolumn{3}{c|}{$\ell$} & \multicolumn{3}{c}{$\ell$} \\
$\ell_1$ & Statistic & $1$  & $2$ & $3$ & $1$  & $2$ & $3$  & $1$  & $2$ & $3$ & $1$  & $2$ & $3$ \\
\hline
\multirow{3}{*}{1} &   $\bar S_{n,1}$   &  \textbf{4.6}  &  2.3  &  1.6 &     \textbf{5.1}   &  2.7  &   2.4 & \textbf{4.5}  & 6.6 &  8.0 &   \textbf{5.6}   &    7.1  &   6.9  &  \cellcolor[gray]{0.85} & \cellcolor[gray]{0.85} &\cellcolor[gray]{0.85}\\
  &       $\bar S_{n,25}$               &  \textbf{4.6}  &  2.3  &  1.6 &     \textbf{5.1}   &  2.7  &   2.4 & \textbf{4.4}  & 6.4 &  7.3 &   \textbf{5.1}   &    6.6  &   7.0  &    \cellcolor[gray]{0.85} & \cellcolor[gray]{0.85} &\cellcolor[gray]{0.85}     \\
  &       $\bar S_{n,50}$               &  \textbf{4.6}  &  2.3  &  1.6 &     \textbf{5.1}   &  2.7  &   2.4 & \textbf{4.5}  & 6.3 &  7.2 &   \textbf{5.2}   &    6.6  &   7.0  &    \cellcolor[gray]{0.85} & \cellcolor[gray]{0.85} &\cellcolor[gray]{0.85}     \\
\hline
\multirow{3}{*}{2} & $\bar S_{n,1}$
&  93.8 & \textbf{7.6} &  3.2   & 100& \textbf{4.0} &  3.1 & 8.1& \textbf{4.3}& 5.6  & 15.8 & \textbf{5.1} & 4.5 &     28.6& \textbf{4.2} &4.8    \\
& $\bar S_{n,25}$                &  93.8   & \textbf{7.6} &  3.2      &   100&    \textbf{4.0}  &   3.1 & 9.2&    \textbf{4.2}&    5.1  & 16.6  &  \textbf{4.8}  &  5.0   &  29.5& \textbf{4.1} &6.0       \\
& $\bar S_{n,50}$               &  93.8   & \textbf{7.6} &  3.2       &   100&    \textbf{4.0}  &   3.1 &9.6&    \textbf{4.3}&    5.4 & 16.3  &  \textbf{4.9}  &  4.8  &   29.4& \textbf{4.1} &5.8       \\
\hline

\end{tabular}
}
\end{table}

\begin{table}[!ht]
\caption{Percentage of rejection of the null hypothesis of $\ell$ regimes at the 5\% level for the DGPs $M_3$ and $M_4$ and for samples of size $n\in\{100,250,500\}$, using $N=1000$ replications and $B=100$ parametric bootstrap samples. Under $\cH_1$, the number of  non-zero regimes is $\ell_1 \in \{1,2\}$, and under $\cH_0$, $\ell=1$ means a non-zero-inflated model. The empirical levels are displayed in bold. }
\label{tab:powerM34}
\centering
{\tiny
\begin{tabular}{cc|ccc|ccc|ccc|ccc|ccc}
\hline
\multicolumn{11}{l}{Exp 1: AR(2) with ${\rm Exp}(1)$ covariate}\\
\hline
& & \multicolumn{3}{c|}{$M_3$, $n=100$} & \multicolumn{3}{c|}{$M_3$, $n=250$} & \multicolumn{3}{c|}{$M_4$, $n=100$} & \multicolumn{3}{c|}{$M_4$, $n=250$} & \multicolumn{3}{c}{$M_4$, $n=500$}  \\
 && \multicolumn{3}{c|}{$\ell$} & \multicolumn{3}{c|}{$\ell$} & \multicolumn{3}{c|}{$\ell$} & \multicolumn{3}{c|}{$\ell$}& \multicolumn{3}{c}{$\ell$}\\
$\ell_1$ & Statistic & $1$  & $2$ & $3$  & $1$  & $2$ & $3$ & $1$  & $2$ & $3$ & $1$  & $2$ & $3$ & $1$  & $2$ & $3$\\
\hline
\multirow{3}{*}{1} & $\bar S_{n,1}$
& 80.2 & \textbf{5.0} & 8.9 & 95.9 & \textbf{5.1} & 3.8 & 13.9 & \textbf{5.2} & 6.0 & 26.0 & \textbf{4.2} & 6.5 & 46.0 & \textbf{4.0} & 5.4\\
& $\bar S_{n,25}$
& 80.2 & \textbf{5.8} & 6.0 & 95.9 & \textbf{5.6} & 2.1 & 22.3 & \textbf{5.2} & 5.5 & 44.4 & \textbf{4.9} & 6.8 & 71.6 & \textbf{4.5} & 5.9 \\
& $\bar S_{n,50}$
& 80.2 & \textbf{4.9} & 6.3 & 95.9 & \textbf{5.3} & 2.2 & 23.4 & \textbf{4.6} & 5.7 & 44.9 & \textbf{6.2} & 6.7 & 71.5 & \textbf{4.0} & 5.4 \\
\hline
\multirow{3}{*}{2} & $\bar S_{n,1}$
& 75.8 & 71.9 & \textbf{3.6} & 98.1 & 97.1 & \textbf{4.4} & 100 & 3.7 & \textbf{3.3} & 100 & 5.4 & \textbf{ 5.4} & 100 &  9.2 & \textbf{4.2 } \\
& $\bar S_{n,25}$
& 75.8 & 78.2 & \textbf{4.0} & 98.1 & 98.3 & \textbf{6.4} & 100 & 3.9 & \textbf{5.0} & 100 & 6.2 & \textbf{ 3.7} & 100 & 12.9 & \textbf{4.4 } \\
& $\bar S_{n,50}$
& 75.8 & 78.5 & \textbf{4.3} & 98.1 & 98.4 & \textbf{6.2} & 100 & 4.2 & \textbf{4.6} & 100 & 6.5 & \textbf{ 4.0} & 100 & 12.2 & \textbf{4.2 }  \\
\hline
\multicolumn{11}{l}{Exp 2: AR(1) with  linear trend}\\
\hline
& & \multicolumn{3}{c|}{$M_3$, $n=100$} & \multicolumn{3}{c|}{$M_3$, $n=250$} & \multicolumn{3}{c|}{$M_4$, $n=100$} & \multicolumn{3}{c|}{$M_4$, $n=250$} & \multicolumn{3}{c}{$M_4$, $n=500$}  \\
 && \multicolumn{3}{c|}{$\ell$} & \multicolumn{3}{c|}{$\ell$} & \multicolumn{3}{c|}{$\ell$}& \multicolumn{3}{c|}{$\ell$} & \multicolumn{3}{c}{$\ell$} \\
$\ell_1$ & Statistic & $1$  & $2$ & $3$  & $1$  & $2$ & $3$ & $1$  & $2$ & $3$ & $1$  & $2$ & $3$ & $1$  & $2$ & $3$\\
\hline
\multirow{3}{*}{1} &   $\bar S_{n,1}$
&98.1 & \textbf{4.8} & 9.6 & 99.9 & \textbf{4.3} & 83.4 & 84.5 & \textbf{4.5} & 6.0 & 99.4 & \textbf{4.1} & 5.9 & \cellcolor[gray]{0.85} & \cellcolor[gray]{0.85} &\cellcolor[gray]{0.85}       \\
  &       $\bar S_{n,25}$
&98.1 & \textbf{5.1} & 6.4 & 99.9 & \textbf{4.2} & 83.4 & 87.2 & \textbf{3.9} & 4.5 & 99.7 & \textbf{2.7} & 6.5 & \cellcolor[gray]{0.85} & \cellcolor[gray]{0.85} &\cellcolor[gray]{0.85} \\
  &       $\bar S_{n,50}$
&98.1 & \textbf{4.5} & 6.1 & 99.9 & \textbf{3.9} & 83.4 & 87.1 & \textbf{4.1} & 4.4 & 99.6 & \textbf{2.6} & 6.2 & \cellcolor[gray]{0.85} & \cellcolor[gray]{0.85} &\cellcolor[gray]{0.85} \\
\hline
\multirow{3}{*}{2} & $\bar S_{n,1}$
&99.9 & 51.2& \textbf{4.2} & 100 & 84.7 & \textbf{5.2} & 100 & 5.8 & \textbf{3.4} & 100 &  6.7 & \textbf{4.3} & 100 &  9.5 & \textbf{4.3} \\
& $\bar S_{n,25}$
&99.9 & 67.0& \textbf{3.4} & 100 & 91.3 & \textbf{3.7} & 100 & 5.4 & \textbf{4.4} & 100 &  7.0 & \textbf{5.5} & 100 & 11.0 & \textbf{4.4} \\
& $\bar S_{n,50}$
&99.9 & 67.3& \textbf{3.4} & 100 & 91.3 & \textbf{3.9} & 100 & 5.8 & \textbf{4.5} & 100 &  7.2 & \textbf{5.2} & 100 & 10.9 & \textbf{4.3}  \\
\hline
\end{tabular}
}
\end{table}

\subsection{Comparisons of the  procedures for the selection of the number of regimes}\label{ssec:selectionExp}
In this section, we investigate the selection of the number of regimes  using the  methodology proposed in Section \ref{ssec:selection}, as well as the classical selection criteria (AIC, BIC, ICL) and the Bayesian inference criteria  (DIC, WAIC, Scott) defined in Section \ref{ssec:information}. First, samples of size $n = 250$ are simulated for the same DGPs. To reduced the computation time, we performed $100$ replications, using $B=100$ parametric bootstrap samples for $S_{n,50}$, while $5000$ iterations were used for the Bayesian methods.
In all cases, we selected the number of regimes $\ell \in \{1,2,3,4\}$. The results of the two experiments for DGPs $M_1$--$M_4$ are presented  in Table \ref{tab:Selection}.

It follows from these results that the Bayesian criteria DIC and WAIC perform rather poorly for regime-switching models, while Scott's method  is correct  70\% (resp. 66\%) of the time for Exp 1 (resp. Exp 2) for Gaussian models, when the true number of regimes is 1,
it is correct 35\% (resp. 50\%) of the time for Exp 1 (resp. Exp 2) when the true number of regimes is 2. For the other models,  Scott's method does not work most of the time, compared to the other non-Bayesian methods.

Next, in terms of performance, for the Gaussian models $M_1$ and $M_3$, the best selection method is BIC, followed closely by ICL and our proposed selection method  based on $\bar S_{n,50}$. There are even some cases where the BIC is less efficient than our method. The AIC criteria behaves rather poorly  compared to $S_{n,50)}$, BIC, and ICL.
For the Poisson models $M_2$ and $M_4$, based on the results in Tables \ref{tab:powerM12}--\ref{tab:powerM34} about power assessment, it is no surprise that when the true number of regimes is $\ell_1=1$, all three classical information criteria perform rather well. This is no longer the case when $\ell_1=2$. In fact, the BIC is the worst criteria, and AIC is the best one, followed by our proposed method.
Finally, in the zero-inflated case $M_4$, our method selects the correct number of regimes $\ell_1=2$ only 41\% of the time for Exp 1, which is coherent with the results in Table \ref{tab:powerM34}, while AIC and BIC are almost perfect. As for Exp 2, our proposed method do as well as the BIC, followed by the AIC. However, when $\ell_1=2$, AIC is the best, followed by our method, and BIC performs rather poorly. To summarize the results, there is no clear winner and one should always compute the AIC, BIC, and our selection method.

Finally, to understand a little better the selection procedure in the Poisson case, we performed a last set of experiments using constant means for each regime. More precisely, we simulated Poisson models with $\ell_1\in \{1,2,3\}$ regimes, $\lambda \in \{1,3,10\}$ and $\lambda\in \{1,6,12\}$, for $n \in \{250,500\}$. The results are displayed in Table \ref{tab:SelecPoisson}. As the number of regime increases, the performance of our proposed method, AIC, and BIC selection methods decreases, and it can be compensated by increasing the sample sizes. When the difference between the means is large enough, the results for these three selection methods are excellent, even for a sample size of $n=250$. Note that ICL criterion and the Baysian methods perform rather badly for selecting the number of regimes in this last experiment.

\begin{table}[!ht]
\caption{Percentage of time a  model with $\ell$ regimes is selected, for the DGPs $M_1--M_4$ for samples of size $n=250$, using for $N = 100$ replications, with $B = 100$ parametric bootstrap samples for $S_{n,50}$, and $5000$ iterations for the Bayesian methods, with a sample size $N_B=100$. Here, $\ell_1$ is the number of non-zero regimes. The correct number of regimes is indicated in bold.}\label{tab:Selection}
\centering
{\tiny
\begin{tabular}{cc|cccc|cccc|cccc|cccc}
\hline
\multicolumn{18}{l}{Exp 1: AR(2) with ${\rm Exp}(1)$ covariate}\\
\hline
&  & \multicolumn{4}{c|}{$M_1$, $n=250$} & \multicolumn{4}{c|}{$M_2$, $n=250$} &  \multicolumn{4}{c|}{$M_3$, $n=250$} & \multicolumn{4}{c}{$M_4$, $n=250$}\\
 && \multicolumn{4}{c|}{$\ell$} & \multicolumn{4}{c}{$\ell$} & \multicolumn{4}{c|}{$\ell$} & \multicolumn{4}{c}{$\ell$} \\
$\ell_1$ & Criteria & $1$  & $2$ & $3$ & $\ge 4$ & $1$  & $2$ & $3$ & $\ge 4$ & $1$  & $2$ & $3$ & $\ge 4$ & $1$  & $2$ & $3$ & $\ge 4$ \\
\hline
\multirow{7}{*}{1}
& $\bar S_{n,50}$         &  \textbf{  91}   &   9   &   0   &    0  &  \textbf{ 98}   &     1   &      0   &    1   &  5   & \textbf{ 88}  &    6  &    1 &    56  & \textbf{41}   &   2   &   1 \\
& AIC                     &  \textbf{  50}   &  14   &  19   &   17  &  \textbf{100}   &     0   &      0   &    0   & 15   & \textbf{ 50}  &   16  &   19 &     0  & \textbf{99}   &   1   &   0 \\
& BIC                     &  \textbf{ 100}   &   0   &   0   &    0  &  \textbf{100}   &     0   &      0   &    0   & 20   & \textbf{ 80}  &    0  &    0 &     1  & \textbf{99}   &   0   &   0\\
& ICL                     &  \textbf{ 100}   &   0   &   0   &    0  &  \textbf{100}   &     0   &      0   &    0   & 20   & \textbf{ 79}  &    1  &    0 &     2  & \textbf{97}   &   1   &   0  \\
& DIC                     &  \textbf{   1}   &  13   &  26   &   60  &  \textbf{ 0 }   &     0   &     12   &   88   &  0   & \textbf{  4}  &   30  &   66 &     5  & \textbf{ 6}   &  23   &  66   \\
& WAIC                    &  \textbf{ 100}   &   0   &   0   &    0  &  \textbf{ 99}   &     1   &      0   &    0   &  0   & \textbf{100}  &    0  &    0 &    52  & \textbf{46}   &   2   &   0 \\
& \text{``Scott''}        &  \textbf{  70}   &  19   &   9   &    2  &  \textbf{ 49}   &    26   &     15   &   10   &  0   & \textbf{  0}  &    0  &  100 &    30  & \textbf{18}   &  22   &  30  \\
\hline
\multirow{7}{*}{2}
& $\bar S_{n,50}$  &      1  &  \textbf{ 98}  &     1 &     0      &     85  &     \textbf{15}    &      0   &    0 &  5  &   0   & \textbf{91}  &    4  &   0  &  93  &  \textbf{  7 }  &   0\\
& AIC              &      0  &  \textbf{ 36}  &    25 &    39      &     65  &     \textbf{35}    &      0   &    0 &  0  &   0   & \textbf{67}  &   33  &   0  &  87  &  \textbf{ 13 }  &   0 \\
& BIC              &      0  &  \textbf{ 95}  &     4 &     1      &    100  &     \textbf{ 0}    &      0   &    0 &  0  &   0   & \textbf{91}  &    9  &   0  & 100  &  \textbf{  0 }  &   0 \\
& ICL              &      0  &  \textbf{ 87}  &     9 &     4      &    100  &     \textbf{ 0}    &      0   &    0 &  0  &   0   & \textbf{85}  &   15  &   0  & 100  &  \textbf{  0 }  &   0 \\
& DIC              &      0  &  \textbf{  2}  &    19 &    79      &       0 &     \textbf{ 0}    &     16   &   84 &  0  &   0   & \textbf{12}  &   88  &   0  &   1  &  \textbf{ 11 }  &  88  \\
& WAIC             &    100  &  \textbf{  0}  &     0 &     0      &     100 &     \textbf{ 0}    &      0   &    0 &  0  & 100   & \textbf{ 0}  &    0  &  84  &  16  &  \textbf{  0 }  &   0\\
& \text{``Scott''} &      8  &  \textbf{ 35}  &    24 &    33      &       0 &     \textbf{ 0}    &      0   &  100 &  0  &   0   & \textbf{ 0}  &  100  &   0  &   9  &  \textbf{ 40 }  &  51 \\
\hline
\multicolumn{18}{l}{Exp 2: AR(1) with linear trend}\\
\hline
&  & \multicolumn{4}{c|}{$M_1$, $n=250$} & \multicolumn{4}{c|}{$M_2$, $n=250$}   & \multicolumn{4}{c|}{$M_3$, $n=250$} & \multicolumn{4}{c}{$M_4$, $n=250$}\\
 && \multicolumn{4}{c|}{$\ell$} & \multicolumn{4}{c}{$\ell$} & \multicolumn{4}{c|}{$\ell$} & \multicolumn{4}{c}{$\ell$} \\
$\ell_1$ & Criteria & $1$  & $2$ & $3$ & $\ge 4$ & $1$  & $2$ & $3$ & $\ge 4$ & $1$  & $2$ & $3$ & $\ge 4$ & $1$  & $2$ & $3$ & $\ge 4$  \\
\hline
\multirow{7}{*}{1}
 & $\bar S_{n,50}$  &  \textbf{  98}   &    2  &    0   &    0    & \textbf{   93}   &     3   &      0   &    4   &   0 & \textbf{ 99}  &    1   &   0  &    3 & \textbf{ 92}  &     2 &    3 \\
& AIC               &  \textbf{  73}   &   14  &    8   &    5    & \textbf{   96}   &     4   &      0   &    0   &   0 & \textbf{ 76}  &   14   &  10  &    0 & \textbf{ 95}  &     5 &    0 \\
& BIC               &  \textbf{ 100}   &    0  &    0   &    0    & \textbf{  100}   &     0   &      0   &    0   &   0 & \textbf{100}  &    0   &   0  &    0 & \textbf{100}  &     0 &    0 \\
& ICL               &  \textbf{ 100}   &    0  &    0   &    0    & \textbf{  100}   &     0   &      0   &    0   &   0 & \textbf{100}  &    0   &   0  &    0 & \textbf{100}  &     0 &    0 \\
& DIC               &  \textbf{   0}   &    0  &    8   &   92    & \textbf{    0}   &     1   &    28    &   71   &   0 & \textbf{  0}  &    1   &  99  &    0 & \textbf{  0}  &    16 &   84   \\
& WAIC              &  \textbf{ 100}   &    0  &    0   &    0    & \textbf{  100}   &     0   &     0    &    0   &   0 & \textbf{100}  &    0   &   0  &   58 & \textbf{ 42}  &     0 &    0  \\
& \text{``Scott''}  &  \textbf{  66}   &   25  &    7   &    2    & \textbf{    0}   &     0   &     0    &  100   &   0 & \textbf{  0}  &    0   & 100  &    0 & \textbf{ 21}  &    29 &   50 \\
\hline
\multirow{7}{*}{2}
& $\bar S_{n,50}$  &       0   &   \textbf{80}   &    19  &    1   &   88  &  \textbf{10}    &     1    &    1  &   0  &  9 & \textbf{ 85}    &    6  &     0 &  89 & \textbf{ 9}  &  2   \\
& AIC              &       0   &   \textbf{35}   &    30  &   35   &   78  &  \textbf{21}    &     1    &    0  &   0  &  0 & \textbf{ 75}    &   25  &     0 &  76 & \textbf{24}  &  0     \\
& BIC              &       0   &   \textbf{62}   &    34  &    4   &  100  &  \textbf{ 0}    &     0    &    0  &   0  &  0 & \textbf{100}    &    0  &     0 & 100 & \textbf{ 0}  &  0   \\
& ICL              &       0   &   \textbf{60}   &    28  &   12   &  100  &  \textbf{ 0}    &     0    &    0  &   0  &  0 & \textbf{ 96}    &    4  &     0 & 100 & \textbf{ 0}  &  0   \\
& DIC              &       0   &   \textbf{ 0}   &     8  &   92   &    0  &  \textbf{ 2}    &    19    &   79  &   0  &  0 & \textbf{  3}    &   97  &     0 &   0 & \textbf{ 1}  & 99   \\
& WAIC             &     100   &   \textbf{ 0}   &     0  &    0   &  100  &  \textbf{ 0}    &     0    &    0  &   0  & 84 & \textbf{ 16}    &    0  &    96 &   4 & \textbf{ 0}  &  0   \\
& \text{``Scott''} &      21   &   \textbf{50}   &    13  &   16   &    0  &  \textbf{ 0}    &     0    &  100  &   0  &  0 & \textbf{  0}    &  100  &     0 &  16 & \textbf{13}  & 71   \\
\hline
\end{tabular}
}
\end{table}

  \begin{table}[!ht]
  \caption{Percentage of time a  model with $\ell$ regimes is selected, for a Poisson model
for samples of size $n\in\{250,500\}$, using for $N = 100$ replications, with $B = 100$ parametric bootstrap samples
for $S_{n,50}$, and $5000$ iterations for the Bayesian methods, with a sample size $N_B=100$. The correct number of regimes is indicated in bold.}
\label{tab:SelecPoisson}
\centering
 {\tiny
 \begin{tabular}{cc|cccc|cccc|cccc}
 \hline
 \multicolumn{14}{l}{$\lambda_1=1$, $\lambda_2=3$, $\lambda_3=10$}\\
 \hline
 &  & \multicolumn{4}{c|}{$\ell_1=1$} & \multicolumn{4}{c|}{$\ell_1=2$} & \multicolumn{4}{c}{$\ell_1=3$}  \\
 && \multicolumn{4}{c|}{$\ell$} & \multicolumn{4}{c|}{$\ell$} & \multicolumn{4}{c}{$\ell$} \\
 $n$ & Criteria & $1$  & $2$ & $3$ & $\ge 4$ & $1$  & $2$ & $3$ & $\ge 4$ & $1$  & $2$ & $3$ & $\ge 4$ \\
 \hline
  \multirow{7}{*}{$250$}
 & $\bar S_{n,50}$    &   \textbf{  97}   &   1  &  1  &  1 &    7  &   \textbf{88}   &      4  &  1  &   0  &   39   &   \textbf{ 54}   &     7   \\
 & AIC                &   \textbf{  98}   &   2  &  0  &  0 &    0  &   \textbf{98}   &      2  &  0  &   0  &    4   &   \textbf{ 87}   &     9   \\
 & BIC                &   \textbf{ 100}   &   0  &  0  &  0 &    1  &   \textbf{99}   &      0  &  0  &   0  &   51   &   \textbf{ 49}   &     0   \\
 & ICL                &   \textbf{ 100}   &   0  &  0  &  0 &   85  &   \textbf{15}   &      0  &  0  &   0  &  100   &   \textbf{  0}   &     0   \\
 & DIC                &   \textbf{   0}   &   1  & 22  & 77 &    0  &   \textbf{ 3}   &     12  & 85  &   0  &    0   &   \textbf{  6}   &    94   \\
 & WAIC               &   \textbf{  96}   &   4  &  0  &  0 &  100  &   \textbf{ 0}   &      0  &  0  & 100  &    0   &   \textbf{  0}   &     0   \\
 & \text{``Scott''}   &   \textbf{  48}   &  21  & 14  & 17 &    0  &   \textbf{26}   &     28  & 46  &   0  &   11   &   \textbf{ 30}   &    59   \\
  \hline
 \multirow{7}{*}{$500$}
 & $\bar S_{n,50}$    &  \textbf{ 93}  &     3 &     0 &   4 &   3 &   \textbf{ 91}  &     3&   3 &   0 &  3 &  \textbf{ 91}  &  6\\
 & AIC                &  \textbf{ 99}  &     1 &     0 &   0 &   0 &   \textbf{100}  &     0&   0 &   0 &  0 &  \textbf{ 93}  &  7\\
 & BIC                &  \textbf{100}  &     0 &     0 &   0 &   0 &   \textbf{100}  &     0&   0 &   0 &  4 &  \textbf{ 96}  &  0\\
 & ICL                &  \textbf{100}  &     0 &     0 &   0 &  83 &   \textbf{ 17}  &     0&   0 &   0 &100 &  \textbf{  0}  &  0\\
 & DIC                &  \textbf{  0}  &     2 &    20 &  78 &   0 &   \textbf{  0}  &   19 &  81 &   0 &  0 &  \textbf{  4}  & 96\\
 & WAIC               &  \textbf{ 97}  &     3 &     0 &   0 & 100 &   \textbf{  0}  &    0 &   0 & 100 &  0 &  \textbf{  0}  &  0\\
 & \text{``Scott''}   &  \textbf{ 69}  &    13 &     8 &  10 &   0 &   \textbf{  0}  &    0 & 100 &   0 &  0 &  \textbf{  0}  &100\\
\hline
\multicolumn{14}{l}{$\lambda_1=1$, $\lambda_2=6$, $\lambda_3=12$}\\
 \hline
 &  & \multicolumn{4}{c|}{$\ell_1=1$} & \multicolumn{4}{c|}{$\ell_1=2$} & \multicolumn{4}{c}{$\ell_1=3$}  \\
 && \multicolumn{4}{c|}{$\ell$} & \multicolumn{4}{c|}{$\ell$} & \multicolumn{4}{c}{$\ell$} \\
 $n$ & Criteria & $1$  & $2$ & $3$ & $\ge 4$ & $1$  & $2$ & $3$ & $\ge 4$ & $1$  & $2$ & $3$ & $\ge 4$ \\
 \hline
 \multirow{7}{*}{$250$}
 & $\bar S_{n,50}$    & \textbf{ 94}   &  2   &   0   &   4  &    1  & \textbf{92}  &   2   &  5    &    0  &   7   &  \textbf{88}   &   5 \\
 & AIC                & \textbf{ 98}   &  2   &   0   &   0  &    0  & \textbf{93}  &   7   &  0    &    0  &   0   &  \textbf{95}   &   5 \\
 & BIC                & \textbf{100}   &  0   &   0   &   0  &    0  & \textbf{98}  &   2   &  0    &    0  &  12   &  \textbf{88}   &   0 \\
 & ICL                & \textbf{100}   &  0   &   0   &   0  &    0  & \textbf{98}  &   2   &  0    &    0  &  97   &  \textbf{ 3}   &   0 \\
 & DIC                & \textbf{  0}   &  3   &  23   &  74  &    0  & \textbf{ 0}  &  15   & 85    &    0  &   0   &  \textbf{ 9}   &  91 \\
 & WAIC               & \textbf{ 98}   &  2   &   0   &   0  &  100  & \textbf{ 0}  &   0   &  0    &  100  &   0   &  \textbf{ 0}   &   0 \\
 & \text{``Scott''}   & \textbf{ 48}   & 20   &  16   &  16  &    2  & \textbf{23}  &  32   & 43    &    0  &   0   &  \textbf{ 3}   &  97 \\
 \hline
\end{tabular} }
  \end{table}

 \section{Example of application}\label{sec:ex}

\begin{figure}[ht!]
     \centering
     \includegraphics[scale=0.75]{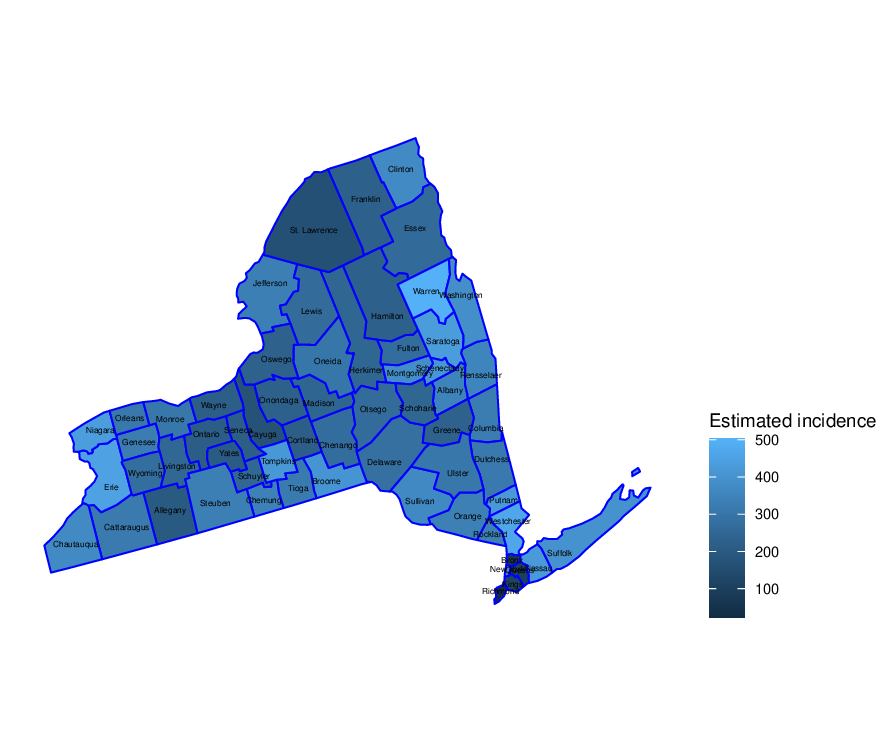}
     \caption{Estimated weekly cases per 100 000 inhabitants (incidence) for New York Counties for May 14 to May 20, 2022.
     Note that the predicted incidence for Bronx, Kings, New York City, Queens, and Richmond, is estimated to be proportional to the predicted incidence of the aggregated New York county. }
     \label{fig:predNY}
 \end{figure}

 To illustrate the proposed methodology, we used the number of reported COVID-19 cases from Jan 1st 2021 to May 20th, 2022, for the 58 counties of New York state reported by CDC. The last seven days are used for comparing predictions. In this example, the variables of interest are  incidence  $y_{tj}$, $j\in \{1,\ldots,58\}$, where $y_{tj}$ is  the daily number of cases by 1000 inhabitants in county $j$.
 Note that for these data, the county of New York is composed of the following counties: Bronx, Kings, New York city, Queens, and Richmond.
 According to the map of New York counties in  Figure \ref{fig:predNY}, we constructed a neighbourhood matrix $M$ with entries $1$ if two counties have common boundaries, and $0$ otherwise. For each county $j\in\{1,\ldots,58\}$,  we chose two explanatory variables:  $Z_{t1,j}\equiv 1$ and $Z_{t,2,j} =\sum_{k=1}^{58} M_{jk} y_{t-1,k}$, with $Z_{1,2,j} = 0$,  i.e., $Z_{t2}$ represents the number of cases by 1000 inhabitants the previous day in the neighbouring counties. The latter variable is used to take into account the spatial dependence. Finally, because there was a significant seasonality of order $7$, we chose to model the seasonal differences $\tilde \by_t = \by_t-\by_{t-7}$. Since we want to predict the last 7 days, for each $j\in\{1,\ldots,58\}$, we estimated the parameters using the first 489 observations $\tilde y_{1,j},\ldots,{\tilde y}_{489,j}$. To this end, we fitted ARX-HMM models with $\ell\in \{2,3,4\}$ and $p\in \{0,\ldots,14\}$ and different distributions. It turned out that the best models were Gaussian. For a given $j\in\{1,\ldots,58\}$, the best model was chosen as the first one for which the $P$-values of the goodness-of-fit test was at least $5\%$, where the models were ordered according to the lexicographic order, i.e., $(2,0),(2,1),\ldots, (2,14),(3,0),\ldots, (4,14)$. The resulting choices are listed in Table \ref{tab:res-covid}.

\begin{figure}[ht!]
 \centering
 \includegraphics[scale=0.35]{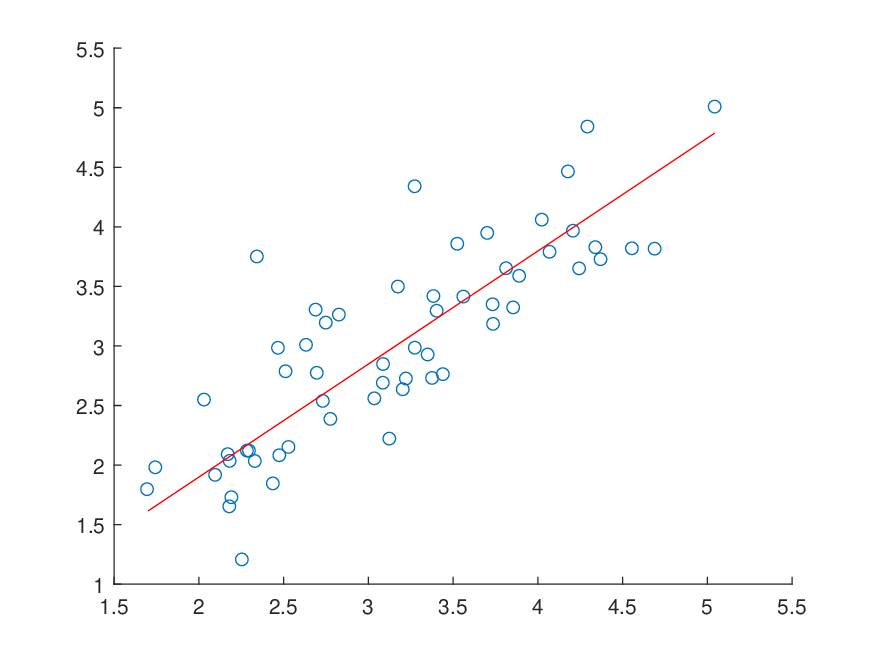}
     \caption{Predicted weekly values vs observed weekly values for all 58 counties. The slope of the regression is $0.94936$.}
     \label{fig:predvsreal}
 \end{figure}

 For the predictions $\widehat{\tilde y}_{t,j}$, since a Gaussian mixture is not symmetric in general, we used the median of the predicting Gaussian mixture using all previous observations.
  The prediction at time $t$ for $y_{t,j}$ is then given by $\widehat{y}_{t,j} = y_{t-7,j}+\widehat{\tilde y}_{t,j}$, $j\in\{1,\ldots,58\}$, and the sum of the predictions were computed to obtain the weekly predictions. As can be seen from Figure \ref{fig:predvsreal}, the results are quite satisfactory, showing that our proposed models make sense. Finally, the severity of the estimated weekly predicted COVID-19 incidence for the New York state counties is illustrated in Figure \ref{fig:predNY}. Note that from the map and Table \ref{tab:res-covid}, the counties with the largest number of predicted weekly incidence are Warren and Weschester.

 \begin{table}[!ht]
 \caption{Best models and predicted weekly incidence for the $58$ counties with $\ell\in\{2,3,4\}$ and $p\in \{0$, $\ldots$, $14\}$. Note the best models were all Gaussian HMMs with this specific number of regimes and lags.}
 \label{tab:res-covid}
  \centering
 {\tiny
 \begin{tabular}{|l|c|c|c|l|c|c|c|} \hline
 County & $\ell$ & $p$ & Pred & County & $\ell$ & $p$ & Pred\\
 \hline
 Albany & 3 & 0      & 356.78    & Niagara & 3 & 2           &  438.18   \\
 Allegany & 2 & 3    & 215.57    & Oneida & 3 & 8            &  301.42     \\
 Broome & 3 & 7      & 407.80    & Onondaga & 2 & 14         &  232.32      \\
 Cattaraugus & 3 & 0 & 294.52    & Ontario & 3 & 7           &  216.85        \\
 Cayuga & 3 & 4      & 174.90    & Orange & 3 & 1            &  334.31   \\
 Chautauqua & 3 & 0  & 371.77    & Orleans & 3 & 0           &  311.41      \\
 Chemung & 2 & 11    & 334.95    & Oswego & 3 & 3            &  235.45    \\
 Chenango & 2 & 2    & 273.19    & Otsego & 3 & 0            &  273.13  \\
 Clinton & 3 & 6     & 374.09    & Putnam & 3 & 6            &  433.15      \\
 Columbia & 3 & 2    & 335.62    & Rensselaer & 2 & 14       &  353.54      \\
 Cortland & 3 & 1    & 217.87    & Rockland & 3 & 2          &  358.75     \\
 Delaware & 2 & 2    & 292.38    & Saratoga & 2 & 11         &  435.93     \\
 Dutchess & 3 & 3    & 319.23    & Schenectady & 4 & 0       &  416.03        \\
 Erie & 3 & 2        & 458.24    & Schoharie & 2 & 11        &  247.33     \\
 Essex & 2 & 12      & 278.28    & Schuyler & 2 & 14         &  256.77     \\
 Franklin & 2 & 9    & 233.41    & Seneca & 2 & 11           &  210.22        \\
 Fulton & 3 & 1      & 277.14    & St. Lawrence & 2 & 7      &  174.61       \\
 Genesee & 2 & 13    & 340.01    & Steuben & 2 & 11          &  336.05       \\
 Greene & 3 & 4      & 244.62    & Suffolk & 3 & 2           &  399.41    \\
 Hamilton & 4 & 1    & 208.74    & Sullivan & 3 & 0          &  366.5     \\
 Herkimer & 2 & 3    & 249.46    & Tioga & 2 & 6             &  329.46        \\
 Jefferson & 2 & 7   & 333.76    & Tompkins & 3 & 0          &  431.06      \\
 Lewis & 2 & 4       & 286.57    & Ulster & 3 & 2            &  309.54    \\
 Livingston & 3 & 4  & 253.58    & Warren & 3 & 0            &  496.74     \\
 Madison & 3 & 0     & 246.44    & Washington & 3 & 1        &  396.98       \\
 Monroe & 3 & 4      & 309.54    & Wayne & 3 & 6             &  231.97     \\
 Montgomery & 3 & 1  & 376.90    & Westchester & 2 & 8       &  470.04        \\
 Nassau & 3 & 0      & 434.94    & Wyoming & 3 & 1           &  274.97     \\
 New York City & 2 &7& 377.56    & Yates & 2 & 13            &  207.82     \\
 \hline
 \end{tabular}
 }
 \end{table}

\section{Conclusion}\label{sec:conclusion}

We showed how to model time series data with ARX-HMM models, i.e., hidden Markov models using previous observations of the series as well as other covariates. Based on the numerical experiments performed in Section \ref{sec:num}, it is proposed that the number of regimes is determined by the first number of regimes for which the $P$-value is at least 5\%, using the averaged Cram\'er-von Mises statistics $S_{n,25}$ or $S_{n,50}$, where the $P$-value is approximated by parametric bootstrap with at least $B=100$ samples. For  choosing the number of regimes, one should also compute the BIC and the AIC (especially for Poisson distribution) for a fixed number of potential candidates for the number of regimes, e.g., for $\ell\in\{1,\ldots,\ell_1\}$.The ICL criterion is either not good or very close to the BIC. This is also true of the Bayesian criteria DIC, WAIC, and Scott.
 An example of an application using COVID-19 cases for New York state counties shows that the proposed models for time series fit the data quite well.

\appendix

\section{EM estimation for general HMM}\label{app:genEM}

\subsection{E-Step}\label{ssec:estepRS}

Using $\disp f_\btheta(\btau,\by,\bz) = \eta_0(\tau_0)\left(\prod_{t=1}^{n} Q_{t,\tau_{t-1},\tau_{t}}\right) \times \prod_{t=1}^n g_{\tau_t,\bbbeta_{\tau_t} }(y_t, \bx_t)$, one has
\begin{equation*}
  \begin{aligned}
   \cQ_{\by,\bz}(\tilde\btheta,\btheta) &= \dE_\btheta\{\log f_{\tilde\btheta}(\btau,\bY,\bZ) |\bY=\by,\bZ=\bz\}\\
     &= \log{\eta_0(\tau_0)}+ \sum_{t=1}^{n} \sum_{k=1}^\ell\sum_{k=1}^\ell \dP_\btheta(\tau_{t-1}=j,\tau_{t}=k|\bY=\by,\bZ=\bz)\log\tilde Q_{t,jk} \\
     & \quad + \sum_{t=1}^n \sum_{k=1}^\ell \dP_\btheta(\tau_t=j|\bY=\by,\bZ=\bz)  \log g_{j,\tilde\bbbeta_j}\left(y_t,\bx_t\right) \\
     &=  \log{\eta_0(\tau_0)}+ \sum_{t=1}^{n} \sum_{k=1}^\ell\sum_{k=1}^\ell \Lambda_{t,\btheta}(j,k)\log\tilde Q_{t,jk}  + \sum_{t=1}^n \sum_{k=1}^\ell  \lambda_{t,\btheta}(j)  \log g_{j,\tilde\bbbeta_j}\left(y_t,\bx_t\right),\\
\end{aligned}
\end{equation*}
where
$
\lambda_{t,\btheta}(j) = \dP_\btheta(\tau_t=j|\bY=\by,\bZ=\bz)$ and $\Lambda_{t,\btheta}(j,k)=  \dP_\btheta(\tau_{t-1}=j,\tau_{t}=k|\bY=\by,\bZ=\bz)$,
for all $t\in \setn$ and $j,k \in \setl$. Now,   for all $j\in\setl$, set $\bar \eta_{n,\btheta}(j)= 1$,   $\eta_{0,\btheta}(j) = \eta_0(j)= \frac{1}{\ell} =\frac{1}{|\cR|}$, and define
$\disp 
\eta_{t,\btheta}(j) = \dP_\btheta(\tau_t = j | Y_1 =y_{1},\ldots,Y_t=y_{t}, \bZ_1=\bz_1,\ldots, \bZ_t=\bz_t)$,  $t=1,\ldots, n$.
Further let $\bar \eta_{t,\btheta}(i)= \bar \eta_{t,\btheta}(i,\by_{t+1:n})$ be the conditional density of $\bY_{t+1:n}$ given $\bY_{1:t}=\by_{1:t}$, $\bZ=\bz$, and $\tau_t=i$.
To compute $\lambda_{t,\btheta}(j)$ and $\Lambda_{t,\btheta}(i,j)$, one can use the following proposition, whose proof is given in Appendix \ref{pf:prop}.

\begin{prop}\label{prop:filt}
For all $j\in\setl$ and $ t=n-1,\ldots, 0$,
\begin{equation}\label{eq:etabar}
\bar \eta_{t,\btheta}(j) =   \sum_{k \in \setl} Q_{t,jk} ~g_{k,\bbbeta_k}( y_{t+1},\bx_{t+1}) \bar \eta_{t+1,\btheta}(k).
\end{equation}
	For all $j\in\setl$ and $t\in \setn$, 	
\begin{eqnarray}\label{eq:eta}
\eta_{t,\btheta}(j) &=& \frac{g_{j,\bbbeta_j}\left(y_t,\bx_t\right)\disp \sum_{i\in \setl} \eta_{t-1,\btheta}{(i)} Q_{t,ij}}{\disp\sum_{i \in \setl} \sum_{k=1}^\ell \eta_{t-1,\btheta}(i) Q_{t,ik} ~ g_{k,\bbbeta_k}\left(y_t,\bx_t\right)} ;\\
\lambda_{t,\btheta}(j) & = & \frac{\eta_{t,\btheta}(j)\bar \eta_{t,\btheta}(j)}{\disp \sum_{k \in \setl} \eta_{t,\btheta}(k)\bar \eta_{t,\btheta}(k)}.\label{eq:lambda}
\end{eqnarray}
For all $i,j\in\setl$ and $ t\in \setn$,
\begin{equation}\label{eq:Lambda}
	\Lambda_{t,\btheta}(i,j) = \frac{\eta_{t-1,\btheta}(i)Q_{t,ij} ~ g_{j,\bbbeta_j}\left(y_t,\bx_t\right)\bar
		\eta_{t,\btheta}(j) }{\disp \sum_{k=1}^\ell\sum_{l\in \setl} \eta_{t-1,\btheta}(k)Q_{t,kl} ~ g_{l,\bbbeta_l}(y_{t},\bx_t)\bar \eta_{t,\btheta}(l)}.
\end{equation}
	In particular, for all $j\in\setl$ and $ t\in \setn$,
	$\disp
	\sum_{k=1}^\ell\Lambda_{t,\btheta}(j,k)=\lambda_{t-1,\btheta}(j)$.
\end{prop}

\begin{rem}\label{rem:etabar}
To avoid numerical problems, one should replace $\bar\eta$ in  \eqref{eq:lambda}--\eqref{eq:Lambda} by its normalized version
\begin{equation}\label{eq:etabarnew}
\bar \eta_{t,\btheta}^\star (j) =  \frac{\bar \eta_{t,\btheta}(j)}{\disp \sum_{i\in \setl} \bar \eta_{t,\btheta} (i)} =  \frac{\disp \sum_{k \in \setl} Q_{t,jk}~ g_{k,\bbbeta_k}( y_{t+1},\bx_{t+1}) \bar \eta_{t+1,\btheta}(k)}{\disp \sum_{i\in \setl}{\sum_{k \in \setl} Q_{t,ik}~g_{k,\bbbeta_k}( y_{t+1},\bx_{t+1}) \bar \eta_{t+1,\btheta}(k)}}.
\end{equation}
Since one divides both numerator and denominator by the same constant, the expressions \eqref{eq:lambda}--\eqref{eq:Lambda}  remain the same.
\end{rem}

\subsection{M-Step}
For this step, given $\btheta^{(k)}$, $\btheta^{(k+1)}$ is defined as
$\btheta^{(k+1) } = \disp \argmax_{\btheta} \cQ_{\by,\bz}\left(\btheta,\btheta^{(k)}\right)$.
Setting $\lambda_{t}^{(k)}(j) =  \lambda_{t,\btheta^{(k)}}(j) $ and $\Lambda_{t}^{(k)}(j,l)=\Lambda_{t,\btheta^{(k)}}(j,l)$,  it follows from Section \ref{ssec:estepRS} that
$$
\btheta^{(k+1) } =  \argmax_{\btheta} \sum_{t=1}^{n} \sum_{j,l\in \setl} \Lambda_{t}^{(k)}(j,l)\log Q_{t,jl} + \sum_{t=1}^n \sum_{k=1}^\ell  \lambda_{t}^{(k)}(j) \log g_{j,\bbbeta_j}\left(y_t,\bx_t\right).
$$
As a result, $\btheta^{(k+1)}  = \left(\bbbeta_1^{(k+1)}, \ldots, \bbbeta_\ell^{(k+1)}, \bkappa_1^{(k+1)},\ldots, \bkappa_\ell^{(k+1)} \right)$, where for each $j\in\cR$,
\begin{eqnarray*}
\bbbeta_j^{(k+1)} &=& \argmax_{\bbbeta_j\in \cB_j} \sum_{t=1}^n  \lambda_{t,\btheta^{(k)}}(j) \log g_{j,\bbbeta_j}\left(y_t,\bx_t\right),\\
\bkappa_j^{(k+1)}  &=& \argmax_{\bkappa_j\in \dR^\ell} \sum_{t=1}^{n} \sum_{l=1}^\ell \Lambda_{t}^{(k)}(j,l)\log{q_{jl}(\bkappa_j,\bx_t)}.
\end{eqnarray*}
Note that in the constant case $\bQ_t =\bQ$ for all $t$, one gets $Q_{jl}^{(k+1)} = \dfrac{\sum_{t=1}^n \Lambda_{t}^{(k)}(j,l) }{\sum_{i=1}^\ell \sum_{t=1}^n \Lambda_{t}^{(k)}(j,i) }$.

\section{Proof of Proposition \ref{prop:filt}}\label{pf:prop}
To simplify notations, parameters are omitted. This proof is an extension of the results in \cite{Caccia/Remillard:2018}.
Let  $t \in \{1,\ldots,n\}$ be given.  Further let $\cG_1$ on $\dR^{t-1}$, $\cG_2$ on $\dR$, and $\cG_3$ on $\dR^{n-t}$  be bounded measurable functions.
It follows from the definition of
conditional expectations that for any $i\in \setl$,
\begin{multline*}
	E\{\cG_1(\bY_{1:t-1}) \cG_2(Y_t)\c\cG_3(\bY_{t+1:n}) \lambda_t(i)|\bZ=\bz\} \\
	= E\{ \cG_1(\bY_{1:t-1}) \cG_2(Y_t)\c\cG_3(\bY_{t+1:n})\I(\tau_t=i)|\bZ=\bz\}
	\\
	\qquad \qquad \qquad = E\left[\cG_1(\bY_{1:t-1}) \cG_2(Y_t) E\left\{\c\cG_3(\bY_{t+1:n}) |\cF_t,\tau_t\right\} \I(\tau_t=i) |\bZ=\bz\right]\\
	= E\left[\cG_1(\bY_{1:t-1}) \cG_2(Y_t) E\left\{\c\cG_3(\bY_{t+1:n}) |\cF_t,\tau_t=i\right\} \eta_t(i) |\bZ=\bz \right].
\end{multline*}
Also,
\begin{multline*}
	E\{ \cG_1(\bY_{1:t-1}) \cG_2(Y_t)\c\cG_3(\bY_{t+1:n})\I(\tau_t=i)|\bZ=\bz\}
	\\
	= \sum_{j_1}\cdots \sum_{j_{t-1}}    \left\{ \prod_{k=1}^{t-1} Q_{ j_{k-1} j_k} \right\} Q_{j_{t-1}i}  \left[ \int  \cG_1(\by_{1:t-1})  \left\{ \prod_{k=1}^{t-1} g_{j_k} (y_k,\bx_k) \right\}
 \fL(dy_1)\cdots \fL(dy_{t-1})\right]  \\
 \times \left\{\int \cG_2(y_t) g_i(y_t,\bx_t)  \fL(dy_t) \right\}\times  \left\{\int \c\cG_3(\by_{t+1:n}) \bar \eta_{t,i}(\by_{t+1:n}) \fL(dy_{t+1})\cdots \fL(dy_{n})\right\}\\
 = \sum_{j_1}\cdots \sum_{j_{t}}    \left\{ \prod_{k=1}^{t} Q_{ j_{k-1} j_k} \right\}   \I(j_t=i)\left[ \int  \cG_1(\by_{1:t-1})\cG_2(y_t) \eta_t(i,\by_{1:t}) \left\{ \prod_{k=1}^{t} g_{j_k} (y_k,\bx_k) \right\}\right.\\
 \qquad \left. \fL(dy_1)\cdots \fL(dy_{t})\right]  \\
 \times   \left\{\int \c\cG_3(\by_{t+1:n}) \bar \eta_{t,i}(\by_{t+1:n}) \fL(dy_{t+1})\cdots \fL(dy_{n})\right\}.
\end{multline*}
This proves \eqref{eq:lambda}.
Similarly, for any $l,i\in \setl$,
\begin{multline*}
	E\{ \cG_1(\bY_{1:t-1}) \cG_2(Y_t)\c\cG_3(\bY_{t+1:n})\I(\tau_{t-1}=l) \I(\tau_t=i)\}
	\\
	= \sum_{j_1}\cdots \sum_{j_{t-1}}   \I(j_{t-1}=l)  \left\{ \prod_{k=1}^{t-1} Q_{ j_{k-1} j_k} \right\} Q_{li}  \left[ \int  \cG_1(\by_{1:t-1})  \left\{ \prod_{k=1}^{t-1} g_{j_k} (y_k,\bx_k) \right\}\right. \\
  \qquad \times \left.  \fL(dy_1)\cdots \fL(dy_{t-1})\right]  \\
 \times \left\{\int \cG_2(y_t) g_i(y_t,\bx_t)  \fL(dy_t) \right\}\times  \left\{\int \c\cG_3(\by_{t+1:n}) \bar \eta_{t,i}(\by_{t+1:n}) \fL(dy_{t+1})\cdots \fL(dy_{n})\right\}\\
 = \sum_{j_1}\cdots \sum_{j_{t-1}}    \left\{ \prod_{k=1}^{t-1} Q_{ j_{k-1} j_k} \right\}  \I(j_{t-1}=l)  \left[ \int  \cG_1(\by_{1:t-1})\cG_2(y_t) \eta_{t-1}(l,\by_{1:t-1}) Q_{li} g_{i} (y_t,\bx_t)\right. \\
  \times \left. \left\{ \prod_{k=1}^{t-1} g_{j_k} (y_k,\bx_k) \right\}
 \fL(dy_1)\cdots \fL(dy_{t})\right]   \times   \left\{\int \c\cG_3(\by_{t+1:n}) \bar \eta_{t,i}(\by_{t+1:n}) \fL(dy_{t+1})\cdots \fL(dy_{n})\right\}.
\end{multline*}
This proves \eqref{eq:Lambda}. As a by-product, one gets \eqref{eq:etabar} and \eqref{eq:eta}, the latter obtained by setting $\c\cG_3\equiv 1$.
 This completes the proof.
\qed

\section{Expression for $\bar S_{n,M}$}\label{app:average-Sn}

Set  $u_{n,t,j} = (1-v_{tj})F_{t,\btheta_n}(y_t-)+  v_{tj} F_{t,\btheta_n}(y_t)$  and define $S_{n,jk} = \int_0^1 \dD_n^{(j)}(u)\dD_{n}^{(k)}(u) du$, $j,k \in\{1,\ldots,M\}$, where
$\dD_{n}^{(j)}(u) = \frac{1}{\sn}\sum_{t=1}^n \{\I(u_{n,t,j}\le u)-u\}$.
Then
$$
\bar S_{n,M} = \int_0^1 \bar \dD_{n,M}^2(u)du = \frac{1}{M^2}\sum_{j=1}^M  \sum_{k=1}^M S_{n,jk} = \frac{1}{M^2}\sum_{j=1}^M S_{n,jj}+ \frac{2}{M^2}\sum_{1\le j<k\le M} S_{n,jk},
$$
where, for any $j,k\in\{1,\ldots,M\}$,
$$
S_{n,jk} = \frac{n}{3}+ \frac{1}{2}\sum_{t=1}^n u_{n,t,j}^2 + \frac{1}{2}\sum_{t=1}^n u_{n,t,k}^2 -\frac{1}{n}\sum_{t=1}^n \sum_{i=1}^n \max(u_{n,t,j},u_{n,i,k}).
$$
Therefore
\begin{equation}\label{eq:Snbar}
\bar S_{n,M}  = \frac{n}{3}+ \frac{1}{M} \sum_{j=1}^M \sum_{t=1}^n u_{n,t,j}^2 - \frac{1}{M^2}\frac{1}{n}\sum_{j=1}^M  \sum_{k=1}^M \sum_{t=1}^n \sum_{i=1}^n \max(u_{n,t,j},u_{n,i,k}).
\end{equation}
Note that $S_{n,jj}$ is also given by formula \eqref{eq:cvm}.

\bibliographystyle{apalike}

\end{document}